\documentclass[a4paper,12pt]{article}
\usepackage{amsfonts}
\usepackage{epsfig}

\newcommand{\clth}{\setcounter{thm}{0}}
\newcommand{\sectionnew}[1]{\section{#1}\clth}

\catcode`\@=11
\def\numberbysection{\@addtoreset{equation}{section}
        \def\theequation{\thesection.\arabic{equation}}}

\newtheorem{thm}{Theorem}[section]
\newcommand{\beq}{\begin{equation}}
\newcommand{\eeq}{\end{equation}}
\newcommand{\beqa}{\begin{eqnarray}}
\newcommand{\eeqa}{\end{eqnarray}}
\newcommand{\nn}{\nonumber \\}
\newtheorem{prop}[thm]{Proposition}
\newcommand {\np}[1]{ {\mathrm{:}}{#1}{\mathrm{:}} } 

\newcommand {\I}[1]{ {\mathrm{Im}}\, {#1} }

\def \a {\underline{\alpha}}

\def \l {\underline{\lambda}}
\def \L {\underline{\Lambda}}
\def \La {\underline{\Lambda}^{(\alpha)}}
\def \Lb {\underline{\Lambda}^{(\beta)} }
\def \s {\sigma}
\def \r {\rho}
\def \e {\underline{\mathrm{e}}}
\def \Q {\underline{Q}}
\def \q {\underline{\mathrm{q}}}
\def \p {\underline{\mathrm{p}}}
\def \ex {\mathrm{e}}
\def \ep {\varepsilon}
\def \h {\underline{h}}
\def \z {\zeta}
\def \b {\underline{\beta}}

\def \A {{\mathcal A} }
\def \d {\delta}
\def \g {\underline{\gamma}}
\def \D {\Delta}
\def \o {\underline{\omega}}

\def \t {\tau}

\def \H {{\mathcal H}}

\def \Z {{\mathbb Z}}
\def \N {{\mathbb N}}
\def \G   {\Gamma}

\def \mod {\, \mathrm{mod}\, }
\def\U1{{\widehat{u(1)}}}
\def \PF {\mathrm{PF}}
\def \IP {\mathrm{IP} }
\def \ch {\mathrm{Ch} }
\def \PFC {\mathcal{PF}_k}

\def \winf{W_{1+\infty}}
\def \nl{\nonumber\\}

\numberbysection

\begin{document}
\begin{titlepage}
\begin{center}

\hfill  \quad DFF 355/09/00 \\
\vspace{.5cm}

{\Large\bf Parafermion Hall States from}

\vspace{.1cm}

{\Large\bf Coset Projections of Abelian Conformal Theories }

\vspace{.6cm}

Andrea CAPPELLI\footnote{E-mail: andrea.cappelli@fi.infn.it.} \\
\normalsize\textit{ I.N.F.N. and Dipartimento di Fisica,} \\
\normalsize\textit{ Largo E. Fermi 2, I-50125 Firenze, Italy}

\vspace{.3cm}

Lachezar S. GEORGIEV\footnote{E-mail: lg@thp.uni-koeln.de.
On leave of absence from I.N.R.N.E., Tsarigradsko Chaussee 72,
BG-1784 Sofia, Bulgaria}\\
\normalsize\textit{Inst. Theoretische Physik, Universitat zu K\"oln,}\\
\normalsize\textit{Z\"ulpicher Str. 77,  50937 K\"oln, Germany }

\vspace{.3cm}

Ivan T. TODOROV\footnote{E-mail: todorov@inrne.bas.bg.}\\
\normalsize\textit{ Institute for Nuclear Research and Nuclear Energy,}\\
\normalsize\textit{ Tsarigradsko Chaussee 72, BG-1784 Sofia, Bulgaria}
\end{center}

\vspace{.3cm}

\begin{abstract}
The $\Z_k$-parafermion Hall state is an incompressible fluid 
of $k$-electron clusters generalizing the Pfaffian state of paired electrons.
Extending our earlier analysis of the Pfaffian,
we introduce two ``parent'' abelian Hall states
which reduce to the parafermion state  
by projecting out some neutral degrees of freedom.
The first abelian state is a generalized $(331)$ state which describes
clustering of $k$ distinguishable electrons and reproduces the
parafermion state upon symmetrization over the electron coordinates.
This description yields simple expressions for the quasi-particle
wave functions of the parafermion state.
The second abelian state is realized by a conformal theory with a 
$(2k-1)$-dimensional chiral charge lattice and it reduces to the 
$\Z_k$-parafermion state {\it via} the coset construction 
$\widehat{su(k)_1}\oplus\widehat{su(k)_1}/\widehat{su(k)_2}$.
The detailed study of this construction provides us
a complete account of the excitations of the parafermion Hall state, 
including the field identifications, the $\Z_k$ symmetry and 
the partition function.
\end{abstract}

\vfill
\hfill September 2000
\end{titlepage}
\pagenumbering{arabic}


\sectionnew{Introduction}

\subsection{Outline of the paper}

Conformal field theories have been successfully applied to
describe the universal properties of quantum Hall states \cite{prange},
such as symmetries, quantum numbers and low-energy dynamics of 
edge excitations  \cite{wen}.
The simplest Laughlin states with filling fraction $\nu=1,1/3,1/5,\dots$
\cite{laugh} are well understood in terms of  abelian conformal theories
with central charge $c=1$; most of our understanding
of edge excitations, including their fractional statistics
 and dynamics, has been drawn from these states.
Furthermore, the theoretical predictions have received important
confirmations by experiments \cite{expe}.

Recent studies have addressed more involved
Hall states, such as that occurring at $\nu=5/2$. 
This {\it plateau} in the second Landau level has no analogue in
the first level (i.e. at $\nu=1/2$); thus, it should be
caused by some new dynamical mechanism.
In Ref.\cite{pfaf}, it was proposed that the electrons form {\it pairs},
in a way similar to BCS pairs in superconductors;
these bosonic pairs can then form a
Laughlin fluid with even denominator filling fraction.
The ($p$-wave) pairing of (spin-polarized) electrons 
is represented in the ground-state wave function 
by the Pfaffian term ${\rm Pf}\left( 1/(z_i -z_j )\right)$,
which involves all possible pairs of electron coordinates.
(Other states with different pairings were also considered \cite{hr}
\cite{wen}.)

There is a well-established relation between Hall
states and conformal field theories, which allows to compute
the filling fractions and the quantum numbers of the excitations;
moreover, the analytic part of electron wave functions correspond to
correlators of conformal fields.
Following Ref.\cite{pfaf}, the Pfaffian term can be reproduced by
the correlator of Majorana fermions in the $c=1/2$ conformal field theory,
i.e. the critical Ising model,
plus the usual $c=1$ boson theory accounting for the charge of excitations.

It is remarkable that the Pfaffian Hall state possesses
excitations with {\it non-abelian} fractional statistics \cite{pfaf}:
the adiabatic transport of one excitation
around another causes a multi-dimensional unitary
transformation within a multiplet of degenerate wave functions,
rather than a sign factor (Fermi statistics)
or a phase (abelian anyon statistics).
The non-abelian statistics is easily understood in the Ising conformal theory:
the spin field $\sigma$ possesses the operator-product expansion
$\sigma \cdot \sigma \sim {\rm Id} + \psi$, with two terms in the
right hand side; therefore, multi-spin correlators expand into several
terms (the conformal blocks), which transform among themselves
under monodromy.
Numerical analyses have shown that the Pfaffian
state has a rather good overlap with the exact ground state
at $\nu=5/2$ \cite{killhr}\footnote{
In contrast with the previous expectations
favouring the Haldane-Rezayi paired state \cite{hr}.};
therefore, there is an exciting possibility that
new phenomena such as electron pairing and non-abelian statistics
could be experimentally observed 
at sufficiently low temperatures for $2 <\nu <3$.

Read and Rezayi have proposed \cite{rr} a generalization of
the Pfaffian to a hierarchy of states in the second
Landau level, which are described by the $\U1\times\PF_k$ conformal theories,
where $\PF_k$ stands for the $\Z_k$ parafermions
\cite{zf}. These parafermion Hall states have filling fractions,
\beq
\label{0.1}
\nu \equiv 2 + \nu_k(M) = 2+ \frac{k}{kM+2}, \quad k=2,3,\ldots
\quad M=1,3,5,\dots \ ,
\eeq
with the Pfaffian corresponding to the case $k=2$ and $M=1$.
The same authors found that these Hall states have good 
overlap with the numerical exact ground states at $\nu=13/5, 8/3$, i.e.
for $M=1$ and $k=3,4$ \cite{rr}.
Furthermore, Hall plateaus at these filling values have been 
experimentally observed by cooling down the sample
at extremely low temperatures \cite{exper}.
Read and Rezayi have shown that the $\Z_k$-parafermion Hall state 
possesses an interesting dynamics: the
Hall fluid is made by clusters of $k$ electrons (generalizing the
pairwise binding for $k=2$) 
and there are excitations with non-abelian statistics; other 
properties were discussed in Ref.\cite{gr}.

This new Hall dynamics is expected to be {\it universal}, i.e.
rather robust under small changes of the microscopic dynamics; 
therefore, it should be possible to describe it directly 
in the effective low-energy conformal field theory.
For example, we would like to find conformal theory arguments
for the quantum numbers of these Hall states and for the
mechanism of clustering which actually yields the $\Z_k$ parafermions 
and non-abelian statistics.

Our general idea is to describe the non-abelian Hall states and
the associated conformal theories by first introducing 
suitable {\it parent} abelian 
theories with same filling fraction and then by projecting down to
the original non-abelian theories\footnote{
See Ref.\cite{abs} for a different attempt to describe non-abelian
states with abelian degrees of freedom.}.
The projection preserves the filling fraction because it only affects
the neutral degrees of freedom; moreover, it fulfills a number of
consistency conditions to be specified later.

This two-step approach can be useful for disentangling the well-understood 
Hall physics of abelian states (such as the Laughlin fluids)
from the new phenomena to be associated with the projection.
The abelian states have a special status in the quantum Hall effect
because they realize the $\winf$ symmetry, the natural symmetry of ``simple''
quantum incompressible fluids under area-preserving reparametrizations
of the plane coordinates \cite{ctz}\cite{w-min}.
The $\Z_k$-parafermion states are only symmetric under the 
sub-algebra $\U1\oplus{\cal W}_k$ \cite{w_k} of the $\winf$ algebra;
therefore, the abelian parent theories provide a framework for  
understanding this remarkable symmetry reduction, and 
the appearance of non-abelian statistics, as being the results of 
the projection.

In Ref. \cite{cgt}, we have already shown that the Pfaffian state
can be described in terms of {\it two} abelian lattice theories 
by projecting out neutral degrees of freedom.
Both parent abelian theories provide interesting 
descriptions of the Pfaffian physics; in particular,
one projection relates the $(331)$ with the Pfaffian states and
describes the corresponding transition from the two-layer 
to the thick-layer $\nu=1/2$ Hall setups which was advocated in Ref.\cite{gww}.

In this paper, we describe the generalization
of these abelian descriptions to the parafermion states 
(see Ref.\cite{esi} for a preliminary account of this work).
In Section 2, we generalize the $(331)$ state
to a $k$-component fluid of distinguishable electrons:
one readily finds that this $c=k$ abelian theory has the
correct filling fraction (\ref{0.1}) and can be describe the
$k$-electron clustering. Actually, $k$ electrons can meet
at the same point, i.e. avoid the Pauli principle,
by taking $k$ different ``colours''.
We show that the conventional Laughlin wave functions of
coloured electrons also describe the 
parafermion Hall states after the anti-symmetrization
over all coordinates, which makes the electrons indistinguishable.
We thus obtain simpler expressions for the known parafermion
wave functions \cite{rr}, as well as new expressions for the
insertion of quasi-particles.

The anti-symmetrization of wave-function coordinates corresponds to 
a projection map between two
conformal field theories with $c=k$ and $c=3k/(k+2)$, respectively; 
besides the $k=2$ case ($c=2 \to 3/2$) discussed before \cite{cgt},
this map has not been analysed in the literature. 
Therefore, this approach is not practical for a detailed description of 
the parafermion conformal theory.
In Sections 3 and 4, we consider the better known description \cite{gn}
of $\Z_k$ parafermions in terms of the coset construction \cite{gko}:
\beqa
\label{1.4}
&&\PF_k=
\frac{\widehat{su(k)_1}\oplus \widehat{su(k)_1}}{\widehat{su(k)_2}} \ ,
\qquad\quad c={2(k-1) \over k+2} \ .
\eeqa
We remark that the numerator is actually a $c=2k-2$ abelian theory;
therefore, the parafermion Hall states can be described by
a $c=2k-1$ abelian state plus the coset projection.
This second parent abelian theory (Section 3) possesses the 
extended affine symmetry 
$\U1\oplus\widehat{su(k)_1}\oplus \widehat{su(k)_1}$;
its charge lattice is {\it maximally symmetric}
in the sense of Ref.\cite{fro}, where it has been denoted by
$(M+2 \; | \; {}^1 A_{k-1} \ {}^1 A_{k-1})$.
The two $su(k)$ symmetries refer to quantum numbers which
can be thought of as representing effective
{\it layer} and {\it iso-spin} indices.

After the coset projection, described in Section 4,
there only remains the $\Z_k$ symmetry relative to the parafermion ``charge'' 
\cite{zf} as it should.
It turns out that this charge is coupled to the fractional (anyon) part of 
the physical electric charge of the quasi-particles, as follows. 
The projection of neutral degrees of freedom in the $c=2k-1$ abelian theory
requires the decoupling of its charged and neutral sectors.
This cannot be done globally within the maximally symmetric lattice,
but give rise to a $\Z_k$ selection rule,
relating the quasi-hole electric charge ($u(1)$ part) to 
the parafermion charge ($\PF_k$ part) modulo $k$.
Therefore, the coset construction of the conformal theories for 
the parafermion Hall states
uniquely determines the complete quantum numbers of the quasi-particle
excitations; namely, it completely defines these Hall states 
starting from the original choice of maximally symmetric lattice.
We explicitly construct the 
super-selection sectors (the irreducible representations) of
these conformal theories, compute their characters  and
write their partition functions \cite{cz}.

The description of quantum Hall states by coset
conformal theories has also been independently proposed in Ref.\cite{fpsw}: 
here, we describe the detailed aspects of this construction for
the parafermion Hall states, after having presented the main points
in the preliminary note \cite{esi}.
In agreement with the authors of Ref.\cite{fpsw}, we believe that
coset projections of abelian theories can be physically relevant,
because the reduction of neutral degrees of freedom and of the 
central charge increases the stability of the Hall state
(criterion (S1) in Ref.\cite{fpsw}).
Let us also quote other projective constructions of non-abelian Hall states 
in the Refs.\cite{schout}, which include the different coset
$\PF_k = \widehat{su(2)_k}/\U1$.

In Section 5, we explicitly work out the coset construction of the 
$Z_3$-parafermion Hall state and analyze the properties of its
quasi-particle excitations, their quantum numbers,
character expressions and fusion rules.
In the Conclusions, we briefly outline a set of
postulates which could characterize rational conformal theories 
describing quantum Hall states, encompassing both abelian 
theories and coset projections.

\subsection{Abelian conformal theories and integer lattices}

We recall that an abelian theory is a rational
conformal field theory with integer central charge $c=n$ and
$\U1^n$ affine symmetry; the current algebra is generated
by $n$ abelian currents $J^a (z)=i\partial_z \phi^a(z)$, extended by
vertex operators $Y(\l,z)=\np{\exp(i\lambda_a\phi^a(z))}$,
such that their ``charges'' $\lambda_a$ are vectors of
a $n$-dimensional lattice $\G$.
For example, the $\widehat{su(k)_1}$ theory is the extension
of $\U1^{k-1}$ theory with the $A_{k-1}$ root lattice. 
In the quantum Hall effect, the lattices satisfy some specific 
physical conditions which are summarized hereafter. 
Following Ref.\cite{fro}, we call
{\it chiral quantum Hall lattice} an odd integral lattice 
associated with a {\it charge} vector $\Q$ of the
dual lattice $\G^*$, which assigns to any lattice point $\q \in \G$ 
the {\it electric charge} $(\q|\Q)$ of the corresponding edge excitation.
In particular, the lattice contains an electron excitation $\q^1$
with unit charge $(\q^1|\Q)=- 1$.
Furthermore, the norm $|\q|^2=(\q|\q)$ of the vectors $\q \in \G$ yields
twice the conformal dimensions, i.e. twice the spin of the corresponding 
excitation, while the scalar product $(\q|\p)$ is the relative
statistics of the excitations $\q$ and $\p$.

The charge vector $\Q$ satisfies the following conditions:
\begin{description}
\item[(i)] 
it is {\it primitive}, i.e., not a multiple of any other vector
$\q^*\in\G^*$;
\item[(ii)] 
it is related to $\nu$ by:
\beq
\label{0.3}
|\Q|^2=\nu;
\eeq
\item[(iii)] it obeys the {\it charge--statistics relation} for boson/fermion
excitations:
\beq\label{0.3b}
(\Q|\q)=|\q|^2 \quad \mod \ 2\ , \qquad {\rm for \ any} \ \q\in\G.
\eeq
\end{description}
{\bf Remark.} 
For Hall states in the second Landau level, $2<\nu \le 3$, such as
those in Eq.(\ref{0.1}),
one should actually replace $\nu \to (\nu-2)$ in (\ref{0.3}), 
because the completely filled first level
is silent but adds two to the filling fraction. 

The non-equivalent quasi-particles in the Hall state correspond to 
the irreducible representations of the extended algebra and are labelled by
the points $\l\in\G^*/\G$ where $\G^*$ is the dual
lattice\footnote{
We recall that $\G^*/ \G$ is a finite abelian group
of order $(\det\G)$ whose multiplication law represents the fusion rules.}.
The low-dimensional chiral Hall lattices have been completely
classified in Ref.\cite{fro}.


\subsection{Projective descriptions of the Pfaffian state}

The $\nu=2+1/2$ Pfaffian Hall state \cite{pfaf} corresponds to the $c=3/2$ 
conformal theory of one $u(1)$ current and the Ising model
(i.e. the Majorana fermion). The two parts describe corresponding
pieces in the ground-state wave function (for even $N$):
\beq
\Psi \left(z_1,\dots,z_N \right) = {\rm Pf}\left( {1\over z_i-z_j} \right)
\ \prod_{1=i<j}^N \ \left( z_i-z_j \right)^2 \ ;
\label{pfaffwf}
\eeq
namely, the $N$-point function of the Majorana fermions produces the Pfaffian
and the vertex-operator correlator yields the usual Laughlin factor.

In Ref.\cite{cgt} we showed that the Pfaffian state
could be realized as a two-step projection of a lattice abelian theory.
The starting point was a $c=3$ theory whose Hall lattice $\G$ is 
{\it maximally symmetric} \cite{fro} (this definition is recalled
in Section 3); its Gram matrix 
$\left( G_\G\right)_{ij}=\left(\q^i|\q^j\right)$ 
has the following form\footnote{
The sign convention for the matrix elements $(G_\G)_{31}=(G_\G)_{13}$
differs from that of Ref.\cite{cgt} as explained in Section 3.}, 
in the basis spanned by the electron charge $\q^1$ and two neutral vectors: 
\beq
\label{G_2}
G_\G =\left[
\begin{array}{rcr}
3 &  \ \ \ 1 & -1 \cr   1 & \; \ \ 2 & 0 \cr  -1 & \;\;\; 0 & 2
\end{array} \right],
\qquad\qquad
Q=\left( 1, 0, 0 \right) \in \G^*   \ .
\eeq
In this equation, we also wrote the components of the
charge vector $\Q$ in the dual basis.
The neutral sub-lattice
of $\G$ (i.e., the one orthogonal to $\Q$) is actually the direct sum of
two root lattices for the $su(2)$ Lie algebra and
therefore the abelian conformal theory possesses the extended
$\U1\oplus\widehat{su(2)_1}\oplus \widehat{su(2)_1}$ affine algebra.

Since the Gram matrix (\ref{G_2}) is not decomposable, the neutral degrees of
freedom are not completely decoupled from the charge; the isospin--charge
separation is necessary for the intended projections of neutral degrees
of freedom and can be achieved in the finer orthogonal lattice $\G_{2,1,1}$: 
the points of the original lattice $\G$ correspond to those points 
in $\G_{2,1,1}$ which satisfy $(-1)^{n_1+n_2+n_3}=1$
where $n_i\in\Z$ are the components in the basis of $\G_{2,1,1}$.
This leads to the $\Z_2$ selection rule, the ``parity rule'', which
couples the independent neutral and charged sectors of $\G_{2,1,1}$: 
\[
\G=\G_{2,1,1} / \Z_2, \qquad\quad G_{\G_{2,1,1}}=
\left[\matrix{2\ 0 \ 0 \cr 0\ 1 \ 0 \cr 0\ 0 \ 1} \right] \ .
\]
Note that the rank-two neutral sub-lattice of $\G_{2,1,1}$ corresponds to
the $c=2$ conformal theory of two Weyl fermions which do not carry
physical charge because the corresponding vectors are orthogonal to $\Q$ 
(\ref{G_2}).

The Majorana fermion describing the Pfaffian state is obtained
by two successive projections \cite{cgt} 
which preserves the filling fraction and the conformal
dimension of the electron field so as to maintain the spin-statistics
relation:
\begin{description}
\item[ \ (i)]
{{\bf Project out the second Weyl fermion} corresponding to the third
basis vector
of $\G_{2,1,1}$. This reduces the Virasoro central charge by $1$ but
preserves the topological
order,i.e., the number of irreducible representations, given by 
the determinant of $G_\G$ \cite{wen}. Taking into account the
$\Z_2$ parity rule (inherited from the maximally symmetric state)
this projection gives rise to the 331 Hall state.
In more standard notations \cite{wen}, corresponding to $Q=(1,1)$,
its Gram matrix is 
$\left( { 3 \ \ 1 \atop 1\ \ 3} \right)$ or more generally
$\left( {m+1 \ \ m-1 \atop m-1\ \ m+1} \right)$ (see Section 2).}
\item[(ii)]
{{\bf Project out the imaginary part of the first  Weyl fermion} 
(i.e., one Majorana fermion) corresponding to  the second basis vector of 
$\G_{2,1,1}$. This neutral  projection
removes  $1/2$ from the central charge and also decreases the topological
order. 
The resulting Pfaffian state is  made by the remaining Majorana fermion, 
the abelian current in the first component and the $\Z_2$ parity rule. }
\end{description}
The latter map between the $(331)$ state and the Pfaffian
could occur in a double-layer Hall device at $\nu=1/2$ \cite{gww},
and represent the transition between a regime of low tunneling
between the layers (distinguishable electrons, $(331)$ state)
to the high-tunneling limit (indistinguishable, i.e. one-species electrons,
Pfaffian state) \cite{cgt}.

The combination of the two projections amounts 
to gauging out the tensor product,
\[
\mathrm{Weyl} \left(\Psi^*,\Psi\right)  \otimes
\mathrm{Ising} \left( \varphi \right)=  \widehat{su(2)_2} \ ,
\]
where the $\widehat{su(2)}$ currents are identified with:
\beq
\label{su2_2}
H(z)=2 \ \np{\Psi^*(z)\Psi(z)}, \quad
E^+(z)= \sqrt{2} \ \varphi(z) \ \Psi^*(z), \quad
E^-(z)= \sqrt{2}  \ \varphi(z) \ \Psi(z) .
\eeq
Finally, taking into account the fact that the neutral sub-algebra is
originally $\widehat{su(2)_1}\oplus \widehat{su(2)_1}$  ,
these two steps can be seen as realizing
 the coset construction (\ref{1.4}) for $k=2$ in the neutral part of
the theory. To combine it to the electric part we use the fact that the full
coset theory inherits the $\Z_2$ parity rule which is preserved by both steps
of the projection.

\setcounter{footnote}{0}


\subsection{The parafermion Hall states and electron clustering} 

The Pfaffian state describes the incompressible Hall fluid of
(spin polarized) electron pairs; actually, the Pfaffian term in the
wave function has been explicitly derived as the long-range limit of a 
BCS-like wave function \cite{green}.
An effective interaction which allows such pairing is given by
a three-body pseudo-potential, for which the Pfaffian
wave function (\ref{pfaffwf}) is the exact zero-energy eigenstate
with lowest angular momentum \cite{pfaf}.

The $\Z_k$-parafermion Hall states introduced in Ref.\cite{rr} 
are the generalizations to incompressible fluids made by  clusters
of $k$ electrons; the corresponding $(k+1)$-body interaction is:
\beq
H = v \sum_{1\le i_1<i_2<\dots < i_{k+1}}^N
\delta^2\left(z_{i_1}-z_{i_2} \right)\ 
\delta^2\left(z_{i_2}-z_{i_3} \right)\ \cdots \  
\delta^2\left(z_{i_k}-z_{i_{k+1}} \right)\ \ .
\label{k1body}
\eeq
The $\Z_k$-parafermion wave functions are zero-energy
states which vanish when $(k+1)$ electrons meet together; the 
ground-state solution of lowest angular momentum:
$(i)$ should not vanish when $k$ electron meet; and $(ii)$ should go to zero
as weakly as possible when a further electron is approaching.
The property $(i)$ indicates that $k$ electrons have formed a
bound state (a cluster), and it is certainly not satisfied by an
ordinary Hall state of (unbound) electrons like the Laughlin states.

As nicely discussed in Ref.\cite{gr}, these two properties imply the
following factorization of the ground-state wave function,
when $k$ coordinates are set equal (say $\{z_1,z_2,\dots,z_k\}$):
\beq
\Psi_k\left(z_k,z_k,\dots,z_k,z_{k+1},\dots,z_N \right) \propto
\prod_{i=k+1}^N \left( z_k -z_i \right)^2\ 
\Psi_k\left(z_{k+1},z_{k+2},\cdots,z_N \right) \ .
\label{factor}
\eeq
The right hand side of this equation contains the same wave function for 
$(N-k)$ electrons. 
The vanishing behaviour $(z_k-z_{k+1})^2$ is the lowest one allowed by
Bose statistics, which is assumed in (\ref{factor});
actually, $\Psi_k$ is the reduced wave function, which
is obtained from the complete one describing the fillings (\ref{0.1}) 
by stripping out a standard antisymmetric factor $\prod_{i>k} (z_k -z_i)^M$,
with $M$ odd. (Equivalently, the reduced wave function (\ref{factor})
describes the (unphysical) bosonic parafermion state with $M=0$).
Therefore, the two properties $(i),(ii)$ of the $k$-cluster Hall fluids 
are accounted for by Eq.(\ref{factor}).

\begin{figure}
\centering
\includegraphics[height=6cm]{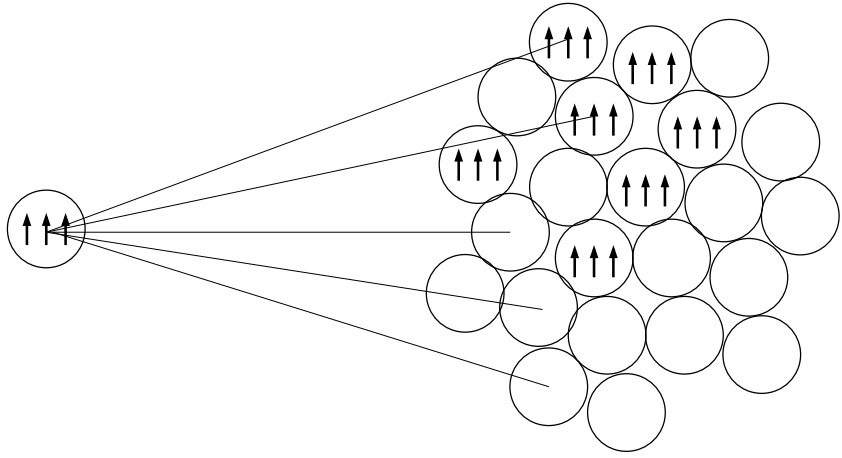}
\caption{Factorization of one cluster in the parafermion Hall state of 
three-electron clusters, as described by Eq.(\ref{factor}).}
\label{fig1}
\end{figure}

The factorization  equation (\ref{factor}) should be interpreted as 
the analogue of the cluster decomposition theorem for statistical models
(see Figure (\ref{fig1}));
in the quantum Hall effect, there cannot be complete factorization
for distant sub-systems, owing to the long-range order: this explains
the factor $\prod_{k<i} (z_k -z_i)^2$.
Another interpretation is that the $k$-electron cluster, once formed
at $z=z_k$, interacts with a further electron at, say, $z=z_{k+1}$, 
as an ordinary particle of a Laughlin fluid.

The Pfaffian wave function (\ref{pfaffwf}) ($M=1,k=2$)  divided by 
$\prod_{k<i} (z_k -z_i)$ does satisfy the factorization equation 
(\ref{factor}). 
For general $k$, the form of $\Psi_k$ is considerably more 
involved and was obtained in Ref.\cite{rr} as follows;
one first introduces a polynomial of $2k$ variables, corresponding to
the two clusters   $\{z_1,\dots,z_k\}$ and $\{z_{k+1},\dots,z_{2k}\}$:
\beqa
\chi_{1,2} \left(z_1,\cdots,z_k;\ z_{k+1},\dots,z_{2k}\right)&=&
(z_1-z_{k+1}) (z_1-z_{k+2})\ (z_2-z_{k+2}) (z_2-z_{k+3})\cdots\nl
&&\cdots (z_k-z_{2k}) (z_{k}-z_{k+1}) \ .
\label{chiwf}
\eeqa
(Note that these factors can be graphically represented by links forming
a closed $2k$-polygon.)
Next, the wave function exists for $N=k n$ electrons, which
are divided in $n $ clusters labelled by the index $r$, $0\leq r<n$,
and reads:
\beqa
\Psi_k\!\!\!\!&&\!\!\!\!\!\!
\left(z_1,\dots,z_k,\dots,z_{kr+1},\dots,z_{kr +k},\dots,z_{(n-1)k+1},
\dots,z_{kn}\right) \nl
&=& {\mathcal S}_N\ \left[\ \prod_{0\leq r < s< n}\ 
\chi_{r,s}\left( z_{kr+1},\dots,z_{kr +k};z_{ks+1},\dots,z_{ks +k}
\right) \right], \quad (N=nk).\ \ 
\label{parawf}
\eeqa
Here, ${\mathcal S}_N$ represents the complete symmetrization over the $N$
coordinates of the expression inside the square brackets, and is properly
normalized to be a projector: ${\cal S}_N^2={\cal S}_N$.

The relation of this wave function to a $N$-point function of the
$\Z_k$-parafermion conformal theory was also found in 
Ref.\cite{rr} and fully proved in Ref.\cite{gr}. It reads:
\beq
\Psi_k \left(z_1,\dots,z_N \right)
\propto \langle \psi_1(z_1)\ \psi_1(z_2)\ \cdots\ \psi_1(z_N)\rangle
       \prod_{1=i<j=N} \left( z_i-z_j \right)^{2/k} \ ,
\label{paracft}
\eeq
where $\psi_1(z)$ is the first of the chiral parafermion fields 
$\psi_\ell(z)$ with conformal dimension $\Delta_\ell$ \cite{zf}:
\beq
\psi_\ell(z)\ ,\qquad \ell\ \mod\ k\ ,\qquad \left(\psi_0 \equiv I \right) ,
\qquad \Delta_\ell= {\ell(k-\ell)\over k} \ .
\label{paraf}
\eeq
In order to prove the equivalence of 
the expressions (\ref{parawf}) and (\ref{paracft}), one first observes that
they are entire functions with equal asymptotic behaviour
(same total angular momentum); then one compares 
their behaviour when $(k+1)$ electrons meet together \cite{gr}. 
In the case of (\ref{paracft}), one can use the operator-product 
expansions \cite{zf}:
\beqa
\psi_\ell(z)\ \psi_{\ell'}(0) &=& z^{-2\ell\ell'/k}\ \psi_{\ell+\ell'}(0)
 \ + \cdots\ , \qquad 0<\ell+\ell'<k \ ;\label{ope1}\\
\psi_\ell(z)\ \psi_{k-\ell}(0) &=& z^{-2\ell(k-\ell)/k}\left(
I\ +\ z^2 \ \ {2\Delta_\ell\over c} T(0) \ + \cdots\right)  \ .
\label{ope2}
\eeqa
Since the fusion of $k$ fields $\psi_1$ yields the identity operator,
the correlator (\ref{paracft}) does not vanish when $k$ electrons meet
at the same point, but becomes the same expression for the remaining
$(N-k)$ coordinates; this checks the factorization formula (\ref{factor}), 
which is also satisfied by the other form (\ref{parawf}) \cite{rr}. 
The identity of the two wave functions is finally proved \cite{gr} 
by checking that there is no linear sub-leading term in the expansion
of (\ref{parawf}) in $(z_{k-1}-z_k)$, at 
$z_1=z_2=\cdots=z_{k-1}$, since this term 
does not appear in operator product (\ref{ope2}) either; such linear term 
would have implied the presence of another $\U1$ current in the
corresponding conformal theory, which is absent in the 
$\Z_k$-parafermion theory \cite{zf}.

The $\Z_k$-parafermion conformal theory also describes
the excitations of this Hall state, which are created by
its primary conformal fields (multiplied by  appropriate choices of 
vertex operators for the $\U1$ part); 
besides the $\psi_\ell$, with abelian fusion
rules and statistics, there exist the spin fields $\s_\ell$ with dimensions
$d_\ell$:
\beq
\s_\ell(z)\ ,\qquad \ell=1,\dots,k-1 \ ,\qquad
d_\ell= {\ell(k-\ell)\over2 k(k+2)} \ .
\label{sigmaf}
\eeq
In particular, the first field $\s_1$ creates the elementary
quasi-hole and obeys non-abelian statistics as in the Pfaffian
case (see Section 5 for examples).

The corresponding wave functions can be written as correlators of 
the spin fields inserted in the electron wave function; 
however, conformal-theory methods do not immediately give explicit
many-particle expressions; 
in Ref.\cite{rr}, the simplest case of $k$ quasi-particle wave function , i.e. 
$\langle\s_1(\eta_1)\cdots\s_1(\eta_k)\psi_1(z_1)\cdots\psi_1(z_N)\rangle$,
was written down.
In the next Section, we shall obtain rather explicit expressions for all
quasi-particle wave functions by using an abelian conformal theory 
and a projection map to the parafermion theory.


\sectionnew{Generalized (331) abelian theory and parafermion wave functions}

\subsection{Ground-state wave function}

The generalization of the $(331)$ abelian theory (recalled in Section
1.2) is a $c=k$ abelian theory with integer $k$-dimensional lattice 
$\tilde \G$ specified by the following Gram matrix and charge vector:
\beqa
\left(G_{\tilde\G}\right)_{ij} &=& \left\{ 
\begin{array}{lll}
M+2 & & i=j=1,\dots, k\ ,\\
M   & & i\neq j\ ,
\end{array} \right.
\qquad M=1,3,\dots,\nl
 \Q &=& (1,1,\dots,1)\ .
\label{kmat}
\eeqa
Form the Gram matrix, also called ``$K$-matrix'' \cite{wen}  in the basis 
(\ref{kmat}), we obtain the following filling fractions,
\beq
\nu_k = \sum_{i,j=1}^k \left(G_{\tilde\G}\right)^{-1}_{ij}=
{k\over kM +2} \ ,
\eeq
which reproduce the values of the parafermion states (\ref{0.1}).
The abelian conformal theory is meant to describe the Hall state of 
$k$ different species of electrons, which are characterized by 
another quantum number conventionally called ``colour''.
There are $k$ different sets of coordinates,
$\{ z^{(a)}_i \}$, $a=1,\dots,k$, because the different species are 
distinguishable and anti-symmetrization is done within each set only.
Each species corresponds to a lattice axis
in the basis (\ref{kmat}), and all species have the same
spectrum of excitations due to the permutation symmetry of $G_{\tilde\G}$.

The ground state wave function is the expectation value of
$k$-component vertex operators, whose conformal dimensions 
can be read off from the $K$ matrix \cite{wen}; the result is,
for $N=nk$ electrons:
\beqa
\!\!\!\!\!\!\!
\Psi_{\tilde\G}\left(
z^{(1)}_1,\dots,z^{(1)}_n;\cdots ;z^{(k)}_1,\dots,z^{(k)}_n
\right) &=& \left[ \prod_{1\le a<b\le k} \prod_{1\le i<j\le n} 
\left( z^{(a)}_i -z^{(b)}_j \right)^M \right]  \nl
&\times & \prod_{a=1}^k \prod_{1\le i<j\le n} 
\left( z^{(a)}_i -z^{(a)}_j \right)^2\ .
\label{331wf}
\eeqa
In order to compare this expression with the reduced ($M=0$) wave function of 
the parafermion Hall states in the previous Section, we shall hereafter
discard the piece between square brackets in the r.h.s of (\ref{331wf}).

It turns out that $\Psi_{\tilde\G}\vert_{M=0}$ satisfies the two
properties for the $k$-electron cluster Hall fluid discussed in the
previous Section: (i) it does not vanish when $k$ electrons meet at 
the same point (provided that they are taken of different colours),
and (ii) it vanishes as $(z_k-z_{k+1})^2$ when a further electron is added.
The first point follows from the fact that distinguishable electrons
do not satisfy the Pauli exclusion principle; the second property is
manifest in (\ref{331wf}).
Note also that the abelian (\ref{331wf}) and the parafermion 
(\ref{paracft}) ground states possess the same total angular momentum.
We conclude that the abelian Hall state (\ref{331wf}) describes
the incompressible fluid of clusters of $k$ {\it distinguishable}
electrons with the filling fractions (\ref{0.1}).

The next step is to describe the case of indistinguishable electrons;
this is achieved by first renaming the coordinates $\{z^{(a)}_i \}$
in (\ref{331wf}) into a single set: 
\beq
{\mathcal R}:\quad
z^{(\ell)}_j \longrightarrow z_{(j-1)k+\ell} \ ,\qquad\quad
j=1,\dots,n, \quad \ell=1,\dots,k,\quad N=kn \ ,
\label{rule}
\eeq
for example $\{z^{(1)}_1,\dots,z^{(k)}_1\}=\{z_1,\dots,z_k\}$,
$\ \ \{z^{(1)}_2,\dots,z^{(k)}_2\}=\{z_{k+1},\dots,z_{2k}\}$;
next, the wave function is symmetrized  over all the $N$ 
coordinates $\{ z_j\}$.
In conclusion, we propose the following ``abelian'' expression for
the $\Z_k$-parafermion ground state (\ref{paracft}):
\beq
\widetilde\Psi_k\left(z_1,\dots,z_N\right) =
\widetilde{\mathcal S}_N \ \left[
 \prod_{a=1}^k \prod_{1\le i<j\le n} \left.
\left( z^{(a)}_i -z^{(a)}_j \right)^2
\right]\right|_{\mathcal R} \ ,\qquad (N=nk)\ ,
\label{abewf}
\eeq
where $\widetilde{\mathcal S}_N=\left(\widetilde {\mathcal S}_N\right)^2$
implements again the full symmetrization and is normalized as a projector.
This expression compares well with the form
(\ref{parawf}): it looks familiar of standard
Laughlin fluids and the physics of $k$-clustering is hidden in the
projection made by $\widetilde{\mathcal S}_N$; let us now prove that
(\ref{abewf}) is another form of the parafermion ground-state
wave function (\ref{paracft}).

{\bf Proof.} We follow the same steps of the previous argument of
the equivalence of the wave functions (\ref{parawf}) and (\ref{paracft}). 
We first argue that (\ref{abewf}) satisfies the factorization 
(\ref{factor}); we rewrite (\ref{abewf}) and consider the limit 
$z_1\sim z_2\sim \cdots\sim z_k$:
\beqa
\widetilde\Psi_k \!\!\!\!\!&&\!\!\!\! \left(z_1,z_2\dots,z_k;
z_{k+1},\dots,z_N\right) \nl
&=& \widetilde{\mathcal S}_N \left. \left[
\prod_{a=1}^k\ \prod_{j=2}^n\left( z^{(a)}_1 - z^{(a)}_j\right)^2 \ 
\prod_{a=1}^k\ \prod_{2\le i<j \le n} \left( z^{(a)}_i -z^{(a)}_j \right)^2 
\right]\right|_{{\mathcal R};\ z_1=z_2=\cdots=z_k} 
\label{abe-clu1}\\
&\propto & \widetilde{\mathcal S}_{N-k} \left. \left[ 
\prod_{a=1}^k\ \prod_{j=2}^n\left( z_k - z^{(a)}_j\right)^2 \ 
\prod_{a=1}^k\ \prod_{2\le i<j\le n} \left( z^{(a)}_i -z^{(a)}_j \right)^2 
\right]\right|_{\mathcal R} \ .
\label{abe-clu2}
\eeqa
We analyse the various terms in the r.h.s. of (\ref{abe-clu1}) 
which are produced by the symmetrization over the coordinates;
there are permutations acting within the first set of $k$ variables 
$I=\{z_1,\dots,z_k\}$ (called $\s'\in{\mathcal S}_k$), those
within the rest of the variables $\bar{I}=\{z_{k+1},\dots,z_N\}$
(called $\s\in{\mathcal S}_{N-k}$), and between the two sets 
($\tau\in{\mathcal S}_{N}$).
The identity permutation identifies $z^{(a)}_1=z_k$, $a=1,\dots,k$,
and yields a non-vanishing contribution, 
because there is no factor $(z_i-z_j)^2$  
connecting pairs of collapsing coordinates
(i.e. there are no ``links'' among them and these coordinates belong 
to a ``cluster'', in the language of Ref.\cite{rr}).
The $\s'$ permutations produce just a multiplicity for the same contribution; 
the $\tau$ permutations exchange one coordinate in $I$
with one coordinate in $\bar{I}$; the form of the wave function is
such that this exchange necessarily introduces
a link between two variables in $I$ and the corresponding
contribution to the wave function vanishes quadratically
at $z_1=\dots=z_k$.

Therefore, all non-vanishing terms of $\widetilde\Psi_k$ in the limit 
$z_1\sim z_2\sim \cdots \sim z_k$ are 
accounted for by the permutations $\s\in{\mathcal S}_{N-k}$, leading to
Eq. (\ref{abe-clu2}); next, we observe that the first factor
in the r.h.s. of (\ref{abe-clu2}) is already symmetric over 
the remaining $(N-k)$ variables,
thus, it can be pulled out of the symmetrization,
and becomes the correct Laughlin factor in the decomposition
(\ref{factor}); the remaining expression inside the square brackets is 
$\widetilde\Psi_k$ for $(N-k)$ electrons, up to an overall factor.
This completes the proof that the wave function (\ref{abewf})
satisfies the decomposition (\ref{factor}).

Next, we proceed to check that the linear sub-leading term in 
the operator-product expansion (\ref{ope2}) is also
absent in the expansion of (\ref{abewf}): following Ref.\cite{gr},
this amounts to show that, after taking $(k-1)$ variables equal,
$z_1=z_2=\cdots=z_{k-1}$, the expansion in $(z_{k-1}-z_k)$
takes the following form to leading order (see Eq.(44) of Ref.\cite{gr}):
\beqa
\widetilde\Psi_k \!\!\!\!\!\!&&\!\!\!\!\!\! \left(
z_{k-1},z_{k-1},\dots,z_{k-1},z_k,z_{k+1},\dots,z_N\right) \nl
&\propto &\left[\left( 
1+ \frac{2(k-1)}{k} (z_{k-1}-z_k) \sum_{i>k}^N {1\over z_k-z_i }
\right)
\prod_{i>k}^N \left( z_k-z_i \right)^2 \right] 
\widetilde\Psi_k \left(z_{k+1},\dots,z_N\right)\ .\ \ 
\label{no-sub}
\eeqa

\begin{figure}
\centering
\includegraphics[height=6cm]{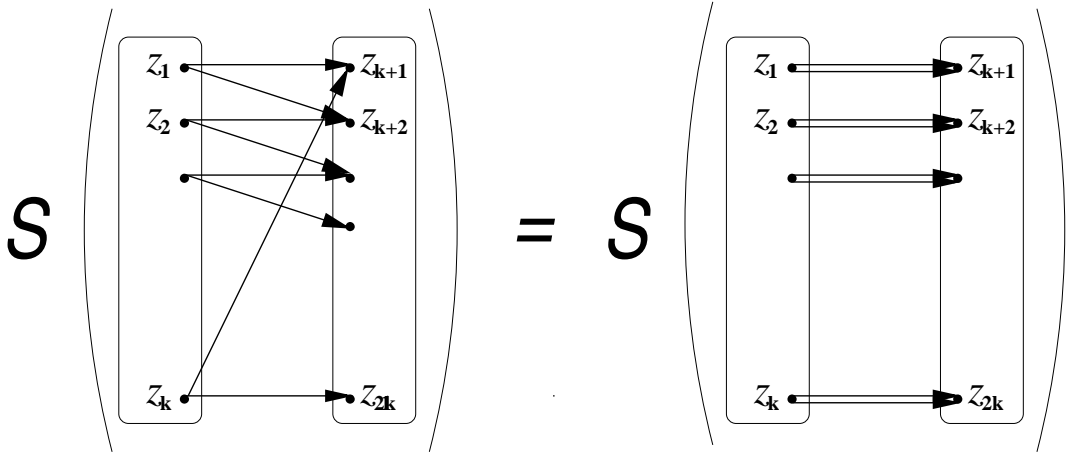}
\caption{Equivalence of clusters interactions for the $\Z_k$-parafermion
wave functions: under coordinate symmetrization, one can 
deviate some of the links $(z_i-z_j)$ (represented by arrows) 
connecting electrons of different clusters.}
\label{fig2}
\end{figure}

In order to analyse the different terms generated by the permutations
in $\widetilde\Psi_k$ (\ref{abewf}), we now divide the coordinates 
into three sets, 
$I=\{z_1,z_2,\dots,z_{k-1}\}$, $\ I'=\{z_k \}$ and 
$\ \bar{I}=\{z_{k+1},\dots,z_N\}$.
Again, the (non-trivial) permutations acting between 
$(I\leftrightarrow \bar{I})$
introduce a link between two coordinates in $I$ and give a 
vanishing contribution to (\ref{no-sub}); the permutations
acting between $(I'\leftrightarrow \bar{I})$ also introduce a link,
which is $(z_{k-1}-z_k)^2$ and thus subleading to (\ref{no-sub}). 
Therefore, one is left with the permutations of
${\mathcal S}_{N-k}$ acting within $\bar{I}$, and the permutations
between $(I\leftrightarrow I')$; a factorization is possible
similar to (\ref{abe-clu2}), with a term corresponding to
$\widetilde\Psi_k(z_{k+1},\dots,z_N)$ and a factor of the form:
\[
\prod_{\{i_1,\dots,i_{k-1},i_k\}} (z_{k-1}-z_{i_1})^2\cdots
(z_{k-1}-z_{i_{k-1}})^2\ (z_k-z_{i_k})^2 \ + {\rm permutations\ of\ }
(i_1,i_2,\dots,i_k) .
\]
The expansion of this expression to order $O(z_{k-1}-z_k)$ matches
the form within square brackets in Eq.(\ref{no-sub});
this completes the proof that (\ref{abewf})
is an equivalent form for the parafermion ground-state wave function 
(\ref{parawf}).
This equivalence can be depicted as in Figure \ref{fig2},
in the case of the two clusters $\{ z_1,\dots,z_k\}$ and 
$\ \{z_{k+1},\dots,z_{2k}\}$ connected by the links $(z_i-z_j)$.

\subsection{Quasi-particle wave functions and the origin of
non-abelian statistics}

The strategy for writing quasi-particle wave functions of 
the parafermion Hall states is now rather clear; we use
standard Laughlin quasi-particle wave functions in the abelian
theory of distinguishable electrons and then apply the symmetrization 
over all coordinates to get the wave functions of the parafermion theory.
From conformal field theory, we know that the 
number of quasi-particle excitations should be a multiple
of $k$ in the $\Z_k$ theory so that the wave functions are
local in the electron coordinates.
For $k$ quasi-holes at positions $\{ \eta_1,\dots,\eta_k\}$, we
can write:
\beqa
\Psi_k \!\!\!\!&&\!\!\!\! \left(\eta_1,\dots,\eta_k;z_1,\dots,z_N\right)= \nl
&& \widetilde{\mathcal S}_N \ \left. \left[
 \prod_{a=1}^k \prod_{i=1}^n \left( \eta_a - z^{(a)}_i \right)\   
\prod_{a=1}^k \prod_{1\le i<j \le n} 
\left( z^{(a)}_i -z^{(a)}_j \right)^2
\right]\right|_{\mathcal R} ,\quad (N=nk) .
\label{k-qh}
\eeqa
This expression can be shown to be equivalent to that 
proposed in Ref.\cite{rr}, by using the same arguments as before.
Note that originally coloured electrons imply coloured quasi-holes,
and thus there is a unique way to write a wave function with
colour matching.

For $mk$ quasi-holes, $m=2,3,\dots$, we can instead write several expressions,
depending on how the quasi-holes are divided into clusters of $k$ members.
For $2k$ quasi-holes, we can e.g. split the coordinates
$\{ \eta_1,\dots,\eta_{2k}\}$
into $(1,2,\dots,k)$ $ (k+1,\dots,2k)$ and obtain the wave function:
\beqa
\Psi_k \!\!\!\!\!\!&&\!\!\!\!\!\! 
\left(\eta_1,\dots,\eta_k;\eta_{k+1},\dots,\eta_{2k};
z_1,\dots,z_N\right) \nl
&=& \widetilde{\mathcal S}_N \ \left. \left[
 \prod_{a=1}^k \prod_{i=1}^n \left( \eta_a - z^{(a)}_i \right)
\left( \eta_{k+a} - z^{(a)}_i \right)\   
\prod_{a=1}^k \prod_{1\le i<j \le n} 
\left( z^{(a)}_i -z^{(a)}_j \right)^2
\right]\right|_{\mathcal R} .
\label{2k-qh}
\eeqa
Analogous expressions can be written for 
the other $2k!/2(k!)^2$ choices of ordering;
however, the number of independent wave functions is 
much lower, because there are non-trivial linear dependences
among these polynomial expressions. 
The actual independent quasi-particle states are better
given by the corresponding conformal blocks of 
the $\Z_k$-parafermion conformal theory. Their number is
computable using the fusion rules of the
fields $\s_1$ with itself and with $\psi_1$: 
for $k=2$, there are $2^{m-1}$ independent $2m$ quasi-hole
wave functions \cite{pfaf}, and for $k=3$ their multiplicities are given 
the Fibonacci number $F_{3m-2}$ \cite{gr}.

These results have been already found for the Pfaffian state in Ref.\cite{nw}:
as an example, we recall the case of the
three possible $4$ quasi-hole wave functions of the Pfaffian state,
with coordinate splittings: $(12)(34)$, $(13)(24)$ and $(14)(23)$
(see Eq. (\ref{2k-qh})); only two of them are
linearly independent due to the identity ($z_{ij}\equiv z_i -z_j$,
$\eta_{ij}\equiv \eta_i -\eta_j$):
\[
z_{12}\ z_{34}\ \eta_{12}\ \eta_{34} +
z_{13}\ z_{42}\ \eta_{13}\ \eta_{42} +
z_{14}\ z_{23}\ \eta_{14}\ \eta_{23} =0\ .
\]

The phenomena of:
i) multiplicities of quasi-particle states for given
positions and quantum numbers, and
ii) linear dependences in the natural basis with simple
monodromy transformations (cf. (\ref{2k-qh})),
have been shown in Ref.\cite{nw}
to characterize the non-abelian statistics of the Pfaffian quasi-holes; 
here, we can extend their analysis to any $k$.
Let us rephrase the argument of Ref.\cite{nw}: the exchange of, say, two
quasi-particles at positions $(\eta_1,\eta_{k+1})$, would map
(\ref{2k-qh}) into the analogous expression for the coordinate splitting: 
$(k+1,2,3,\dots,k)$ $(1,k+2,k+3,\dots,2k)$;
if all the wave functions $(i_1,\dots,i_k)$
$(i_{k+1},\dots,i_{2k})$ were linear independent, these exchange
transformations could be simultaneously diagonalized and would amount to
multiplications by a phase, i.e. to abelian anyon statistics.
However, the wave functions form a degenerate basis, such that the
exchanges are represented by non-trivial (non-diagonalizable) matrices
on the basis of independent functions, leading to
non-abelian statistics.

These two options are exemplified by the quasi-particle wave functions 
of the parafermion Hall state (\ref{2k-qh}) 
and of the parent abelian state (namely, Eq.(\ref{2k-qh}) without 
$\widetilde{\mathcal S}_N$), respectively:
\beqa
no\ projection &\to& independent\ basis\ \to\ \ \ abelian\ statistics, \nl 
projection\ \ \ \ \ &\to& degenerate\ basis\ \ \to\ \ non-abelian\ statistics. 
\label{2cases}
\eeqa

In conclusion, we have shown that the origin of non-abelian statistics
can be simply and quite generally understood as the effect of the 
projection from a parent abelian Hall state; without the reference 
to an abelian state, the  non-abelian statistics
would have seemed to be the consequence of rather mysterious 
algebraic properties of specific wave functions \cite{nw}.
The other side of the story is that the projection also breaks the
generic $\winf$ symmetry of abelian Hall states \cite{ctz} down to the
${\cal W}_k$ symmetry of $\Z_k$ parafermions \cite{w_k}; 
it would be interesting to develop a physical picture for this symmetry
reduction and its interplay with non-abelian statistics
(this point will be further discussed in Section 4.5).
 
\bigskip

More work (mostly numerical analyses) needs to be done for extracting 
physical information from the quasi-hole wave functions (\ref{2k-qh}).
As for exact results, using conformal methods, the parent abelian
theory presented in this Section is not particularly convenient,
because the projection to the parafermion theory is a non-standard mapping
between conformal theories with central charges $c=k$ and $c=3k/(k+2)$,
respectively.
Therefore, it is useful to consider another parent
abelian theory with $c=2k-1$: although larger, this theory admits
a known mapping to the parafermions, the coset
construction (\ref{1.4}) \cite{w_k}, which yields the
complete conformal-theory description of the parafermion
quasi-particle excitations.


\sectionnew{Maximally symmetric parent abelian theory for the para\-fermion 
Hall states}


\subsection{Maximally symmetric $(2k-1)$-dimensional lattice}

In this Section, we describe the abelian theory
with central charge $c=2k-1$ and affine symmetry 
$\U1\oplus\widehat{ su(k)_1}\oplus\widehat{ su(k)_1}$, which also reproduces
the filling fractions (\ref{0.1}) of the parafermion theory.
It is described by a {\it maximally symmetric} Hall lattice in 
the terminology of Ref. \cite{fro}.  
We recall that abelian theories with maximally symmetric lattices
reproduce almost all observed filling fractions, including the 
Jain plateaus \cite{fro};
however, the present theory corresponds to different ranges of filling 
fraction $\nu_k$, i.e. the {\it sub-windows} \cite{fro}:
\beq\label{0.1b}
\Sigma_p^-=
\left\{ \nu \ ; \frac{1}{2p}\leq \nu < \frac{1}{2p-1}\right\}
\quad for  \;\; M=2p-1 \ .
\eeq
As noted in \cite{fro}, the properties of Hall states in these sub-windows
are strikingly different from the better-known complementary ones with
$\nu\in \Sigma_p^+=\left[\frac{1}{2p+1}, \frac{1}{2p} \right)$,
that include the Laughlin series $\nu=1/3, 1/5, \ldots$.

We shall briefly recall the notion of  {\it maximal symmetry} \cite{fro}.
A chiral  Hall lattice  is called
{\it maximally symmetric} if its neutral sub lattice ($\G_0 \perp \Q$)
coincides with the (internal symmetry) Witt sub-lattice
 $\G_W \subset \G$  (defined as the sub-lattice generated by
all vectors of square length 1 or 2) so that dim $\G_W=$ dim $\G  - 1$.
The Witt sub-lattice is always
of the type $A\oplus D \oplus E \oplus \Z$.
The maximally symmetric lattices, denoted in \cite{fro} by the
symbol $(L| \, {}^{\o}\G_W)$, can be characterized in an appropriate basis
 by the  following Gram matrix:
\beq
\label{MSQHL}
G =\left[
\begin{array}{c|c}
L &  \o \cr \hline
\o^T & C(\G_W)
\end{array}
\right] \ ;
\eeq
the charge vector is $Q=(1,0, \ldots, 0)$ in the corresponding dual basis.
Here $L\in\N$ is the {\it minimal relative angular 
momentum}\footnote{
In this case, it is equal to twice the conformal dimension of 
the electron operator, i.e. an odd integer.} 
that appears as
the minimal power of $(z_i - z_j)$ in the corresponding wave function;
$C(\G_W)$ is the {\it Cartan matrix} for the Witt sub-lattice and
$\o$ is an {\it admissible weight} for $\G_W$ restricted by the
condition  $(\o | \o)<L$ (see Eq. (5.6) in \cite{fro}).
Note that the Witt sub-lattice does not contribute to the filling
fraction $\nu$ since it is orthogonal to the charge vector $\Q$.
It is easy to find $\nu$ for this lattice using (\ref{0.3}),
\beq
\label{nu-msqhl}
\nu = (\Q|\Q) = Q^T. G_\G^{-1}.Q=\frac{1}{L-(\o|\o) }\ .
\eeq

The next step is to derive the maximally symmetric Hall lattice which
reproduces the filling fraction $\nu_k(M)$ in (\ref{0.1}); 
we combine Eqs. (\ref{nu-msqhl}) and (\ref{0.1}) so that:
\[
\frac{1}{L-(\o|\o) } = \frac{1}{M+\frac{2}{k}}=
\frac{1}{M+2 -(2-\frac{2}{k})} \ .
\]
We identify  $L\equiv M+2$ since $L$ should have the same parity as $M$.
In order to find $\o$, we recall that the square of the
$i$-th $su(k)$-fundamental weight  is  $i(k-i)/k$.
Therefore, we can write:
\beq\label{omega}
\o=\La_1 \pm \Lb_1, \quad
(\o|\o)= (\La_1|\La_1) + (\Lb_1|\Lb_1) = 2\frac{k-1}{k}\ ,
\eeq
where
$(\alpha)$ and $(\beta)$ denote two commuting copies of $su(k)$ with roots
$\a^i$ and $\b^j$ and fundamental weights $\La_i$ and $\Lb_j$
respectively, i.e., $(\La_i|\a^j)=(\Lb_i|\b^j)=\d_i^j$, and,
\beq\label{2.4}
(\a^i|\a^j)=(\b^i|\b^j)= C^{ij}, \quad
\quad (\La_i|\La_j)=(\Lb_i|\Lb_j)=
C^{-1}_{ij}=\frac{i(k-j)}{k}, \quad i\leq j, \quad
\eeq
where $C^{ij}$ is the $su(k)$ Cartan matrix 
(all other inner products vanish).
This fixes the Witt sub-lattice to be $\G_W= A_{k-1} \oplus A_{k-1} $.

A convenient basis $\{ \q^a \}_{a=1}^{2k-1}$ for $\G$ is given by the 
electron vector $\q$ (satisfying (\ref{0.3}) and the spin-statistics
relation (\ref{0.3b})) and the root vectors
of $su(k)\oplus su(k)$:
\beqa\label{q-basis}
&&\q^1 \;\;  = \q\ , \nn
&&\q^{1+i}=\a^i \ , \qquad\qquad i=1, \ldots , k-1\ , \nn
&&\q^{k+i}=\b^i \ , \qquad\qquad i=1, \ldots , k-1\ .
\eeqa
One important consequence  of the specific  structure of the Gram matrix
(\ref{MSQHL}) is the following decomposition of the electron vector $ \q $:
\beq
\label{1.3}
\q= - \frac{1}{\nu_k}\ \Q +\o \ , 
\eeq
where  $(\Q|\o)=0$ and $(\a_i|\o)=\d_{i1}=-(\b_i|\o) $.
Note that the present choice of sign for $\o$ slightly differs
from the one in Ref.\cite{cgt}, but it fits
the standard notation for coset projections (see Section 4).

The Gram matrix in the basis (\ref{q-basis}) takes the form:
\beq\label{gram}
\left[G_\G^{ab} \right] =\left[\left( \q^a| \q^b \right)\right]=
\left[
\begin{array}{c|c|c}
M+2 &  1 \; 0  \cdots  \ 0 & -1 \;  0  \cdots \ 0\cr \hline
\begin{array}{c}
1 \cr 0 \cr \vdots \cr 0
\end{array} &  C_{k-1} & 0 \cr \hline
\begin{array}{c}
- 1\cr 0 \cr \vdots \cr 0
\end{array} &  0 & C_{k-1}
\end{array} \right]\ ,
\eeq
with $C_{k-1}$ the Cartan matrix of the $A_{k-1}$ algebra 
(recall that the charge vector is $Q=(1,0,\ldots,0)$ in the dual basis).
This completes the derivation of the maximally symmetric abelian
lattice theory with $c=2k-1$ and filling fractions (\ref{0.1});
its edge excitations are described by representations
of the associated chiral algebra 
$\U1\oplus\widehat{ su(k)_1}\oplus\widehat{ su(k)_1}$, which will be
discussed in the next Section.

We now describe the decomposition of the lattice (\ref{q-basis}), (\ref{gram})
into neutral and charged sub-sectors for the later purpose of
projecting out neutral degrees of freedom and obtaining
the parafermion conformal theory (Section 4).
The strategy is the following:
we should first identify the abelian representations with zero charge
and the corresponding points in $\G$; next,
we single out a sub-lattice $\G$ which is
the direct sum of a one-dimensional charge sub-lattice and its orthogonal
complement; finally, 
we only project the neutral points which are orthogonal to the electron
vector (\ref{1.3}) (and also orthogonal to $\o$).
The last condition implies that the conformal dimension
of the electron, i.e. the spin-statistics relation, is preserved
by the projection: 
\beq\label{dim}
\D_{\mathrm{el}}=\frac{1}{2} (\q|\q)=\frac{1}{2} \left(\frac{1}{\nu_k} + 
(\o|\o) \right) =\frac{M+2}{2} \ ;
\eeq
(note that $\o$ yields a crucial contribution according to Eq. (\ref{1.3})).

The unitary representations of the chiral algebra $\A(\G)$ 
(describing the edge excitations)
are labelled by the points of the dual lattice
$\G^*$: this is manifestly not decomposable into orthogonal
sub-lattices of charge and neutral excitations.
Nevertheless, this decomposition (an instance of the isospin-charge
separation) can be achieved at the expenses of enlarging the 
dual lattice (of excitations)
and of introducing a selection rule (the $\Z_k$ {\it parity rule}).
We introduce the decomposable sub-lattice 
$L\subset \G$ of index $k$ spanned by the vectors:
\beq\label{Lbasis}
\{ \e^1, \a_i, \b_j \}\ , \qquad \e^1 = k(\q-\o)=(kM+2)\Q \ .
\eeq
It splits
into 3 mutually orthogonal sub-lattices:
\beq\label{1.5}
L=(kM+2)\Z\Q \oplus A_{k-1}\oplus A_{k-1} \ .
\eeq
We have
\beqa\label{1.6}
L\subset \G\subset \G^* \subset L^*, &\quad&
\G=\{ \g=\l+n\ \q \; ; \; \l\in L, \; 0\leq n \leq k-1 \}, \nn
&& L^*/\G^* \simeq \G/L \simeq \Z_k;
\eeqa
indeed, the determinants of the Gram matrices of $L$ and $\G$
(which give the number of sectors of the corresponding rational conformal theory) are:
\beq\label{1.7}
|L|=(kM+2)^2 |\Q|^2 |C_{k-1}|^2=(kM+2)k^3 =k^2 |\G|.
\eeq
The isospin-charge separation of excitations is achieved
in the decomposable dual lattice $L^*$, whose physical points
(corresponding to the points of $\G^*$) obey a selection rule $mod \ k$.
This is the desired factorization of excitations in the abelian theory
which will allow the projection of neutral degrees of freedom.


\subsection{Unitary irreducible representations of the chiral 
       algebra $\A(\G)$. The $\Z_k$ parity rule}

The irreducible unitary representations of the
chiral algebra $\A(L)$ can be expressed as
$\Z_k$-invariant products of fundamental representations of
$\widehat{su(k)_1}\oplus \widehat{su(k)_1}$  times chiral vertex
operators carrying charge $(n/k)\Q$, with $n\in\Z/k(kM+2)\Z$;
these representations are labelled by the elements of the abelian group
$L^*/L$. We shall choose a vector $\l\in L^*$ in each coset in $L^*/L$
of the form:
\beq
\label{1.8}
\l=m\ \e^*_1 +\L_\mu^{(\alpha)} - \L_\rho^{(\beta)},
\qquad \mu,\rho=0,1,\ldots, k-1, \qquad 2|m|\leq k(kM+2).
\eeq
Here  $\L_\mu^{(\alpha)} $ ($\L_\rho^{(\beta)}$) are the fundamental
weights (including 0) of the first (respectively, the second) $su(k)$
factor. The conformal dimension of the representation $\l$ is:
\beq
\label{1.9}
\D(\l)=\frac{1}{2} |\l|^2 = \frac{m^2}{2k(kM+2)} + \frac{\mu(k-\mu)+\rho(k-\rho)}{2k}.
\eeq

The charges $\g^*\in \G^*/ \G$ representing the excitations of the
chiral algebra $\A(\G)$ could be viewed as points in $L^* /L$
since we have $\G^*/ \G \subset L^* /L$ by Eq.(\ref{1.6});
this is convenient because of the spin--charge
decomposition of $L$ and $L^*$.
However, we need a criterion to select the points of  the original dual lattice
$\G^*$ among those of the bigger lattice $L^*$: since $\q$ is the only vector
in the basis of $\G$ which does not belong to $L$,
we find the following natural condition.
\begin{prop}
A vector $\l$ of $L^*$ belongs to the sub-lattice $\G^*$ if and only if
$(\l|\q)\in \Z$. For $\l$ given by (\ref{1.8}), this is equivalent
to the relation:
\beq
\label{1.10}
m+k \left(\o|\L_\mu^{(\alpha)}-\L_\rho^{(\beta)} \right) \in k \Z \qquad 
\Longleftrightarrow
\qquad \mu + \rho=m \ \ \mod k.
\eeq
\end{prop}
{\bf Proof.}
Since $\q\in \G$ and $\l\in\G^*$ the inner product $(\l|\q)$ is
integer. For $\l\in L^*$ the converse is also true in view of (\ref{1.3}) and
(\ref{1.5}). Eq. (\ref{1.10}) is then a consequence of (\ref{0.3})
(\ref{1.3}) and (\ref{1.5}) and of the definition of fundamental weights
as a dual basis for $ \{ \a_i \} $ and $ \{ \b_i \} $.
As a result, the inner products $(\L_i|\L_j)$ (for $\L$ belonging
to the weight space of the same $su(k)$ factor, $\alpha$ or $\beta$)
are expressed in terms of the inverse $A_{k-1}$ Cartan matrix.

Equation (\ref{1.10}) is the explicit form of the $\Z_k$ parity rule which
selects the points in $L^*$ that also
belong to the sub-lattice $\G^*$ of physical excitations of the
Hall state.

According to Eq. (\ref{1.6}) the charges labelling the irreducible
representations of $\A(\G)$ can be written in the basis of $L^*$
as follows, taking into account (\ref{1.8}) and (\ref{1.10}):
\beqa
\label{g-charges}
\g^*=\l + n\q &=& m\ \e^*_1 +\La_{\rho+m} - \Lb_{k-\rho} +
n\left( \frac{k}{\nu_k}\e^*_1 + \La_1 - \Lb_1\right) \nn
&=& \left(m+n(kM+2) \right) \e^*_1 + \La_{\rho+n+m} - \Lb_{n-\rho}\ ,
\eeqa
where we have used the fact that the sum of $su(k)$ fundamental weights
$\L_\mu +\L_\rho $ belongs to the sector of $\L_{\mu+\rho \mod k}$.
Therefore, we can label the irreducible representations of $\A(\G )$
by a pair $(m,\rho)$
corresponding to Eq. (\ref{g-charges}), where $m$ measures the minimal
charge of each irreducible representation
so that $2|m| \leq (kM + 2) $, while $\rho \mod k$
characterizes the neutral part.

Using the representation of each vector $\g\in \G$ as a sum
of an $\l\in L$ and a {\it gluing vector} \cite{fro} $n \q\in \G$
(see (\ref{1.6})), we can write the representation space
$\H^\G_{\l}$ of $\A(\G)$ as direct sums of representation spaces of $\A(L)$,
\beq\label{spaces}
 \H^\G_{\l} = \bigoplus_{n=0}^{k-1} \H^L_{\l+n\q} \ ,
\eeq
where $\l\in L^*/L$ is supposed to satisfy the parity rule (\ref{1.10}),
so that $\l \in \G^*/\G$.
This form of $\H^\G_{\l}$ allows to write the corresponding characters
$\chi^\G_{\l}(\t,\z) =
\mathrm{tr}_{\H^\G_{\l}} \; q^{L_0- c/24} e^{2\pi i(\Q|\underline{J_0})}$
as sums of characters of $\A(L)$ modules:
\beq\label{chars_sum}
\chi^\G_{\l}(\t,\z) = \sum_{n=0}^{k-1} \chi^L_{\l+n\q}(\t,k\z)\ ,
\eeq
where the factor $k$ in the argument $k\z$ comes from the normalization 
of the charge vector in the basis of $L^*$.
This expression is convenient since the character  $\chi^L_{\l+n\q}(\t,k\z)$
of $\A(L)$ is simply the product of a one-dimensional lattice character
\cite{cgt} (corresponding to the charged sub-lattice):
\beq\label{1.17}
K_l(\t,\zeta;p)=\frac{1}{\eta(\t)} \sum_{n\in\Z}
q^{\frac{p}{2}(n+\frac{l}{p})^2} \ex^{2\pi i \zeta (n+\frac{l}{p})},
\eeq
and two copies of  the (restricted, i.e., neutral)
$\widehat{su(k)_1}$ characters $\chi_\rho(\t)$
\cite{kt},
\beq\label{1.16}
\chi_\rho(\t,A_{k-1})=\frac{1}{(\eta(\t))^{k-1}}
\sum_{\g\in A_{k-1}} q^{\frac{1}{2}|\L_\rho +\g|^2}, \quad q=\ex^{2\pi i \t},
\quad \rho=0,1,\ldots, k-1 \ .
\eeq
($\eta$ is the Dedekind function and $A_{k-1}$ is the root lattice).
Therefore,  we can write the characters $\chi_{\l}(\t,\zeta)$
of the representation $\l$ (\ref{1.8}) 
as  sums of factorized $L$-characters,
\beq\label{1.18}
\chi_{m, \rho}^\G(\t,\zeta)=
\sum_{n=0}^{k-1} K_{m +n(kM+2)}\left(\t,k\zeta;k(kM+2)\right)
\chi_{\rho+m+n}^{(\alpha)}(\t)   \chi_{n-\rho}^{(\beta)}(\t),
\eeq
with $m$ mod $(kM+2)$ and $\rho$ $ \mod k $.
Note that, originally $m$ was an index defined mod $k(kM+2)$; however,
Eq. (\ref{1.18}) shows that the characters are $\Z_k$ invariant,
$\chi_{m+(kM+2), \rho}^\G(\t,\zeta)=\chi_{m, \rho}^\G(\t,\zeta)$,
so that there are just $(kM+2)$ independent values for $m$.
The resulting set of $k(kM+2)$ functions is covariant under (weak)
modular transformations generated by
$T^2: (\t,\zeta)\rightarrow (\t+2,\zeta)$ and
$S:(\t,\z) \rightarrow (-1/\t,\zeta/\t)$, which are the modular
properties suitable for quantum Hall systems \cite{cz}.


\sectionnew{The $\Z_k$-parafermion coset and its representations}

In this Section, we describe the coset construction (\ref{1.4});
this can be done in each of the $k$ sectors (\ref{spaces}) of the abelian
theory according to the $\Z_k$ parity rule.


\subsection{$\Z_k$ selection rule for triples of su(k) weights.
Conformal dimensions of coset representations}

The $\PF_k$ coset module (\ref{1.4}) is labelled, in principle, by a triple
$(\L_{\alpha},\L_{\beta};\L)$; the pair $(\L_{\alpha},\L_{\beta})$ of
fundamental $su(k)$ weights and the level 2 weight $\L$ fix an irreducible
unitary representation
of the numerator and of the denominator current algebra, respectively.
The tensor product of irreducible $\widehat{su(k)_1}$-modules corresponding
to the numerator in the right hand side of Eq. (\ref{1.4}) splits into a
direct sum of $\PF_k$ and $\widehat{su(k)_2}$-modules:
\beq\label{3.1}
\H^{(1)}_{\L_{\alpha}} \otimes  \H^{(1)}_{\L_{\beta}} = \bigoplus\limits_{\L}
\H(\L_{\alpha},\L_{\beta};\L)\otimes \H^{(2)}_{\L} \qquad
\left( \alpha,\beta=0,\ldots,k-1, \; \L_0=0\right).
\eeq
Not all triples $(\L_{\alpha},\L_{\beta};\L)$  are admissible (i.e.,
correspond to non-empty coset modules) and different
{\it admissible triples} may refer to {\it equivalent representations}.
The following statement is a
specialization of results on field identification (based on the use of simple
currents) obtained in Ref.\cite{schw}.
\begin{prop}
Admissible triples are characterized by the conservation of the $\Z_k$
charge given by the $k$-ality:
\beq\label{3.2}
[\L]=\sum_{i=1}^{k-1} i \ \lambda_i \quad
\mathrm{for} \quad \L=  \sum_{i=1}^{k-1}  \lambda_i \L_i ;
\eeq
more precisely, the triple $(\L_{\alpha},\L_{\beta};\L)$  is
admissible if and only if:
\beq\label{3.3}
 [\L_{\alpha}]+[\L_{\beta}]=[\L] \ \ \mod k , \qquad
\mathrm{i.e.,} \quad \alpha +\beta =[\L] \ \ \mod k.
\eeq
There are thus $k {k+1 \choose 2}$ admissible triples of the form
$(\L_{\alpha},\L_{\beta};\L_{\alpha +\kappa}+\L_{\beta -\kappa})$
where all indices are taken $\mod k$. They split into $ {k+1 \choose 2}$
families of equivalent triples of the form:
\beq\label{3.4}
(\L_{\alpha+\s},\L_{\beta+\s};\L_{\alpha +\kappa+\s}+\L_{\beta -\kappa+\s}) ,
\quad \s=0,\ldots, k-1.
\eeq
As a result, the number $N(\PF_k)$ of parafermion coset sectors coincides
with the number of unitary irreducible representation  $N(\widehat{su(k)_2})$
 of the level 2 current algebra:
$ N(\PF_k)=N(\widehat{su(k)_2})= k(k+1)/2$.
\end{prop}
We can define a representative for each family in Eq.(\ref{3.4})
by choosing a value for $\s$: we set
$\beta+\s=0 \mod k$, thus normalizing the
second fundamental weight to zero. 
Therefore, the equivalent classes of triples can be labelled by
the level-two weight, say $\L_\mu+\L_\r$, $\mu\leq \r$; we have:
\beq\label{3.6}
(\L_{\mu+\r \mod k}, 0;\L_\mu+\L_\r ) \quad \Longleftrightarrow\quad
\L_\mu+\L_\r \quad (\mu\leq \r).
\eeq
Ultimately, these labels characterize the parafermion representations.
We end up with the following characterization of $\PF_k$ coset modules which
appears to be new.
\begin{prop}
The parafermion coset modules are in one to one correspondence with
sums $\left(\L_\mu+\L_\r \right)$ of $su(k)$ fundamental weights
($0\leq \mu \leq \r \leq k-1$). Their conformal weights are given by:
\beqa
\label{3.7}
\D^{(k)}_{\mu\r}&=&\frac{1}{2} \left| \L_\mu +\L_\r \right|^2 -
\D_2\left( \L_\mu +\L_\r \right) \nn
& = &
\frac{\mu(k-\r)}{k}+\frac{(\r-\mu)(k+\mu-\r)}{2k(k+2)}
\quad \mathrm{for} \quad \left(0\leq\mu\leq\r<k \right),
\eeqa
where $\D_2(\L)$ is the dimension of the representation of the level 2
weight $\L$ of $\widehat{su(k)_2}$. The decomposition of a product of level-$1$
characters corresponding to the tensor product expansion (\ref{3.1}) 
for any triple (\ref{3.6}) has the form:
\beq\label{3.8}
\chi^{(1)}(\L_{\a-\b})   \chi^{(1)}(\L_0) =
\sum_{\gamma} \ch\left(\L_{\alpha-\beta +\gamma} + \L_{k-\gamma}\right)
\chi^{(2)} \left(\L_{\alpha-\beta +\gamma} + \L_{k-\gamma}\right) \ ,
\eeq
where $\ch(\L)$ is the coset character and $\gamma$ takes
all values corresponding to distinct sums
of pairs of weights (they are no more than  $(k/2)+1$). 
\end{prop}
{\bf Proof.}
Eq. (\ref{3.7}) can be verified by using the identity
$ \D_1(\L_\mu) + \D_1(\L_{k-\r}) -   \D_2(\L_{\mu+k-\r})  =
\D^{(k)}_{\mu\r} $
and observing that the triple $(\L_\mu,\L_{k-\r};\L_{\mu+k-\r})$
is equivalent to  $(\L_{\mu+\r},0;\L_{\mu}+\L_{\r})$.
The triples appearing in Eq. (\ref{3.8}) (as arguments of the pair of
$\chi^{(1)}$ and $\chi^{(2)}$) are, clearly, admissible. It is
straightforward to verify that the difference of conformal  weights
of the two sides is an integer.
The number of terms in the expansion (\ref{3.8}) is independent
of $\alpha$  and $\beta$. For the vacuum representation we have, for example:
\[
\chi^{(1)}(\L_0)   \chi^{(1)}(\L_0) =  \ch(2\L_0) \chi^{(2)}(2\L_0)+
\sum_{\gamma=1}^{\IP(\frac{k}{2})}  \ch\left(\L_{\gamma} + \L_{k-\gamma}\right)
\chi^{(2)} \left(\L_{\gamma} + \L_{k-\gamma}\right) \ ,
\]
where $\D^{(k)}_{\gamma, k-\gamma}+ \D_2(\L_\gamma +\L_{k-\gamma})=\gamma$
for $\gamma\leq k-\gamma$,
and $\IP(x)$ stands for the integer part of the real number $x$.


\subsection{Parafermion Hall states. The parafermion $\Z_k$ charge}

The chiral algebra $\A_k$ of the $\Z_k$-parafermion Hall states, 
with filling fraction (\ref{0.1}), is determined from:
\beq\label{3.11}
\A_k \otimes \A(\widehat{su(k)_2})=\A(\G).
\eeq
This is obtained by coset projection from the lattice
theory of Section 3. In particular, the lattice
characters (\ref{1.18}) are projected into:
\beq
\label{3.12}
\chi_{m \r } (\tau, \z) = \sum_{s \mod  k} K_{m+ s(kM+2)}
(\tau , k\z; k(kM+2))\ \ch(\tau, \L_{s+m+\r} +\L_{s-\r} ) .
\eeq

The coset projection only preserves a single $\Z_k$ symmetry of 
the original product $\Z_k \times \Z_k$ of the
centres of the two $su(k)$ groups; this is:
\beq
\label{3.13}
\left[\L_{\mu} + \L_{\r} \right]= \mu +\r \qquad \mod\ k \ , 
\eeq
which defines the $\Z_k$ charge of parafermions 
$p=\mu+\r$.

Moreover, the $\Z_k$ parity rule of the parafermion Hall states
is inherited from the corresponding selection rule of the parent
abelian state, given by Eq.(\ref{1.8}): 
this requires that the physical Hall excitations possess
parafermion ``charge'' (\ref{3.13}) 
equal ($mod \ k$) to the number of ``fractional units'' of electric charge 
$l \in \Z_{kM+2}$ in Eq.(\ref{1.8}) (see also Eq.(\ref{1.10})):
\beq\label{p-rule}
p=\mu+\r =m \quad {\rm mod}\ k\ .
\eeq

The coset representations $2\L_\r$, corresponding to $\mu=\r$,
are the {\it ``parafermion currents"} of Fateev and Zamolodchikov
\cite{zf}. They are ``simple currents'' \cite{schw}
obeying $\Z_k$ fusion rules:
\beq\label{3.14}
2\L_{\mu} \times 2\L_{\r}  \sim   2\L_{(\mu+\r) \mod k} .
\eeq
The (non-local) parafermion currents give rise
to an ``anyonic" chiral algebra, say, $\PFC$
whose bosonic (integer dimension fields') sub-algebra can be
 identified with the coset chiral algebra $\PF_k$ (\ref{1.4}). 
The parafermion algebra $\PFC$ admits $k$ unitary irreducible representations, 
labelled by an integer  $\s \mod  k$, with conformal weights,
\beq\label{3.15}
\D_\s = \frac{\s(k-\s)}{2k(k+2)}, \quad \s= 0,1,...,k-1.
\eeq
Each of these splits into $(k-\s)$ unitary irreducible representation 
of the bosonic sub-algebra
$\PF_k$ whose conformal weights exceed (\ref{3.15}) by an integer multiple
of $1/k$.  Comparing (\ref{3.15}) with (\ref{3.7})
we see that $\s$ can be identified  with $(\r-\mu)$. For each $\s$
in the range  (\ref{3.15}) there are  exactly $(k-\s)$ pairs $(\mu,\r)$
satisfying $0\leq \mu\leq\r\leq k$, $\s=\r-\mu$; they generate all
different conformal weights (\ref{3.7}). Following Ref. \cite{schil}, the
characters of the resulting coset modules are given by
$\ch(\tau,\L_\mu+\L_\r)= \ch_{\s l}(\tau)$, where:
\beq\label{3.16}
\ch_{\s l}(\tau)=q^{\D_\s-\frac{c_k-1}{24}}
\sum\limits_{\underline{n}\in \N_l}
\frac{q^{\underline{n}\cdot C^{-1}\cdot (\underline{n}- \L_{\s} ) }}
{ (q)_{n_1}\cdots (q)_{n_{k-1}}},
\quad l\geq \s.
\eeq
In this equation, $\D_\s$ is the  $\PF_k$ weight (\ref{3.15}), $(c_k -1)$
is the parafermion central charge (cf. (\ref{1.4})),
$ (q)_n=\prod_{j=1}^n (1-q^j)$,
$\N_l=\left\{ \underline{n}=(n_1,\ldots,n_{k-1})\right.$; 
$n_i \in \Z_+ ;$
$ \left. n_1+2n_2+\cdots + (k-1)n_{k-1}=l \mod k\right\}$, and
$C^{-1}$ is the inverse of the $su(k)$ Cartan matrix.

The expression (\ref{3.16}) corresponds to a $\PF_k$ irreducible
component of the representation $\s$ of the non-local parafermion
algebra $\PFC$. For fixed $\s$,
the values of $l$ yielding inequivalent $\PF_k$ modules are:
\beq\label{3.21}
l = \s, \s +1,\ldots,k-1 \;\qquad (\mathrm{for} \; \s =0,1, \ldots, k-1).
\eeq
The pair $(\s,l)$ is related to $(\mu,\r)$ of (\ref{3.7}) by:
\beq\label{3.22}
\s = \r - \mu\ ,\ l=\r \quad {\rm i.e.} \quad \mu = l-\s\ ,\ \r =l \quad
\Rightarrow\quad 2l-\s \equiv m \mod  k.
\eeq
The {\it topological order} $N_k$  of the $\Z_k$-parafermion Hall state
is equal \cite{cz} to the number of independent characters (\ref{3.12})
of the corresponding conformal theory; 
it is given by the product of the range of $m$, i.e. ($kM + 2$), 
multiplied by the number ${ k+1 \choose 2}$ of $\Z_k$-parafermion
sectors and divided by $k$, due to the parity rule (\ref{p-rule}):
\beq\label{3.23}
N_k = \frac{1}{k}(kM + 2) { k +1 \choose 2} = \frac{k+1}{2}(kM + 2 ) \ .
\eeq
We thus recover the value obtained in Ref.\cite{rr} by other means.


\subsection{Quasi-particle excitations of parafermion Hall states}

The charge and conformal dimensions of all particles in the theory can be
deduced from their characters -- note that
the characters (\ref{1.18}) and (\ref{3.12}) are represented as sums of product
of ``charged" and ``neutral" characters that always satisfy the 
$\Z_k$ parity rule.
In the coset, the charge and total conformal dimension of each term
$ K_{m+s(kM+2)}(\t,k\z;k(kM+2) ) \ \ch(\tau, \L_{s+m+\r} + \L_{s-\r }) $
in Eq. (\ref{3.12}) are given by:
\beqa\label{q.n}
&&Q=k \frac{m+s(kM+2)}{k(kM+2)} = \frac{m+s(kM+2)}{kM+2} \ , \nn
&& \D_{\mathrm{tot}} = \frac{(m+s(kM+2))^2}{2k(kM+2)} +
\D^{(k)}(\L_{l+\s + s} + \L_{s-\s }) \ .
\eeqa
The physical hole operator (whose charge conjugate is the
electron operator) is identified with the tensor product:
\beq\label{hole}
\Psi_{\mathrm{hole}}(z) =
\np{\ex^{i\frac{1}{\sqrt{\nu}} \phi(z)}} \ \psi_1(z) =
\np{\ex^{i\frac{kM+2}{\sqrt{k(kM+2)}} \phi(z)}} \ \psi_1(z) \ ,
\eeq
satisfying the parity rule (\ref{p-rule})
($\psi_1$ being the $l=1$ parafermion field in Eq.(\ref{paraf})). 
The sector corresponding to one hole is given by the $s=1$ term in 
the character,
\beq\label{vacuum}
\chi_{00}(\t,\z)= \sum_{s=0}^{k-1} K_{s(kM+2)}(\t,k\z;k(kM+2) ) \
\ch(\L_{s} + \L_{s})\ ,
\eeq
so that its charge and conformal dimension are:
\[
 Q_{\mathrm{hole}} = 1,
\qquad
\D_{\mathrm{hole}} = \frac{1}{2\nu_k} + \frac{k-1}{k} =
\frac{M+2}{2} \ .
\]
Note that the conformal dimension of the physical electron (or hole)
is independent of $k$.

Similarly the quasi-hole operator is identified with:
\beq\label{q.h.}
 \Psi_{\mathrm{q.h.}}(z) =
\np{\ex^{i\frac{1}{\sqrt{k(kM+2)}} \phi(z)}} \ \s_1(z)\ ,
\eeq
where $\s_1$ is the spin field with the label $(\L_0 +\L_1)$
corresponding to  the lowest charge and dimension among the
irreducible representations of the coset (see Eq.(\ref{sigmaf})). 
It is also worth noting that 
it generates all other irreducible representations by fusion with itself.
It is characterized by the $s=0$ term in
\beq\label{q.h.char}
\chi_{10}(\t,\z)= \sum_{s=0}^{k-1} K_{1+s(kM+2)}(t,k\z;k(kM+2)) \
\ch(\tau, \L_s +\L_{s+1})\ ,
\eeq
 so that the charge and total conformal dimensions are:
\beq\label{q.h.q.n.}
 Q_{\mathrm{q.h.}} = \frac{1}{kM+2},
\quad
\D_{\mathrm{q.h.}} = \frac{1}{2k(kM+2)} + \frac{k-1}{2k(k+2)} 
\stackrel{(M=1)}{=}\frac{1}{2(k+2)} \ .
\eeq


\subsection{Coset description of parafermion fusion rules}

All fusions rules can be obtained from those of the basic fields
$\L_0+\L_\mu$, $0\leq \mu \leq k-1$ (see Eq.(\ref{3.6})),
taking into account the symmetry of the fusion coefficients under 
the charge conjugation and the action of the simple currents:
\beq\label{fusion}
(\L_0+ \L_\mu) \cdot (\L_0+\L_\rho) =
\left\{ \begin{array}{cl}
\quad \bigoplus\limits_{s=0}^{\min(\mu,\rho)}  \quad (\L_s+\L_{\mu+\rho-s}) 
&\mathrm{if} \quad  \min(\mu,\rho)\leq \IP(k/2) \ ,\\
& \\
\bigoplus\limits_{s=k-\max(\mu,\rho)}^{k}   (\L_s+\L_{\mu+\rho-s})
& \mathrm{otherwise}\ .
\end{array} \right.
\eeq
Note that the second case $\min(\mu,\rho) > \IP(k/2)$ in 
Eq. (\ref{fusion}) is 
obtained from the first one
by using the charge conjugation symmetry $C$ of the fusion coefficients
$ N_{C(\l),C(\underline{\mu})}^{C(\underline{\rho})} =
 N_{\l,\underline{\mu}}^{\underline{\rho}}$ where  
$\underline{\lambda},\underline{\mu},\underline{\rho}$ are labels 
in the form of Eq. (\ref{3.6}) and 
$C(\L_\mu+\L_\r)=(\L_{k-\mu}+\L_{k-\r})$.

The other fusions could be obtained by the action of the simple currents,
\[
  N_{J(\l),J'(\underline{\mu})}^{J.J'(\underline{\rho})} =
 N_{\l,\underline{\mu}}^{\underline{\rho}} \ ,
\]
where $J$ and $J'$ are arbitrary powers (up to $k$) of the generator $I$ 
of the simple currents orbit:
\[
I(\L_\mu+\L_\r)=(\L_{\mu+1}+\L_{\r+1}) .
\]
For example, applying $I\times I$ to both sides of Eq. (\ref{fusion}) 
(for $\mu \leq \IP(k/2)$ ) we get: 
\beqa
(\L_1+\L_{\mu+1}) \cdot (\L_1+\L_{\r+1})&=&
I(\L_0+\L_\mu) \cdot I(\L_0+\L_\r)   =
\bigoplus\limits_{s=0}^{\min(\mu,\rho)}  I^2 (\L_{s}+\L_{\mu+\rho-s})
\nl
&=& \bigoplus\limits_{s=0}^{\min(\mu,\rho)}  (\L_{s+2}+\L_{\mu+\rho-s+2}).
\eeqa

A complete list of fusion rules for the coset irreducible representations 
 will be given in Section 5 for the $k=3$ case.


\subsection{${\cal W}_k$ symmetry of the parafermions: the spectrum
of ${\cal W}_k$ charges in the 
  coset $\widehat{su(k)_1}\oplus \widehat{su(k)_1} / \widehat{su(k)_2} $.}

A characteristic property of the parafermion coset (\ref{1.4})
is its preservation of the chiral algebra ${\cal W}_k$ generated
by quasi-primary fields of dimensions $2, \ldots, k$  corresponding
to the $su(k)$ polynomial Casimir operators
(the quadratic one being the Virasoro stress tensor); instead, the 
dimension-one currents of the denominator $\widehat{su(k)_2} $ are 
all gauged away. Altogether, 
the symmetry $\widehat{su(k)_1}\oplus \widehat{su(k)_1}$ 
of the parent abelian theory is reduced to the Casimir algebra ${\cal W}_k$.
This result follows from the fact that the conformal fields corresponding to
all the Casimir operators commute with the action of the simple currents.
Comparison of Eq. (\ref{1.4}) with the formula
for the Virasoro central charge of the ${\cal W}_k$-symmetric minimal 
conformal theories \cite{w_k},
\beq
c_p\left({\cal W}_k  \right) = (k-1) \left( 1- \frac{k(k+1)}{p(p+1)}\right)
, \quad p=k+1, k+2, \ldots\ ,
\label{c-w}\eeq
indicates that the $\Z_k$-parafermion theories correspond to the lowest models
in each series, i.e. $p=k+1$.

The occurrence of the ${\cal W}_k$ symmetry in the parafermion Hall
states is rather intriguing because it also characterizes the better-known
Jain Hall states \cite{prange}: according to the proposal of Ref.\cite{w-min},
the Jain states are described by conformal theories corresponding
to the highest models in the series
(\ref{c-w}), at the limiting value $p=\infty$, which have been
called  ``$\winf$ minimal models''.
Both realizations of the ${\cal W}_k$ symmetry come from projections of
abelian theories, which are symmetric
under the larger algebra $\winf$ ($\U1\oplus{\cal W}_k\subset \winf$).
The latter algebra describes the deformations of planar incompressible
fluids under area-preserving diffeomorphisms of the plane; thus, it accounts 
for the dynamical symmetry of ``simple'' incompressible Hall fluids 
\cite{ctz}.

The projection leading to the $\winf$ minimal models is rather mild,
it reduces the multiplicities of the excitations but 
does not change the central charge \cite{w-min}: 
the physical picture associated to the
basic incompressible Hall fluids, such as the Laughlin fluids,
is preserved \cite{ctz}.
On the other hand, the projection leading to the parafermion theories
is much stronger and the physics of the resulting Hall fluids is
rather different from that of the Jain states (and of generic abelian states),
as discussed in the previous Sections.

In spite of the differences, the appearance of the ${\cal W}_k$ symmetry
in both cases suggests the possibility of a simple physical mechanism
underlying both stable Hall states.
For this reason, it is interesting to briefly describe
the characterization of the $\Z_k$-parafermion sectors as irreducible
representations of the ${\cal W}_k$ algebra; one needs
the eigenvalues of the zero modes of all $(k-1)$  ${\cal W}_k$ ``currents"
(the first of them, the stress tensor, giving the conformal dimension)
to label the lowest-weight states.
Of course, in Section 4.1 we introduced the simpler labelling by 
two quantum numbers: $\mu$ and $\rho$ $(\mu\leq \rho)$ or $\sigma$ and $l$.
(The second choice is more convenient as it gives the $\Z_k$ charge
and the label $l-\sigma$ of the simple current (of weight $2\L_{l-\sigma}$)
acting on the lowest weight state $\L_0+\L_\sigma$.)

The properties of the ${\cal W}_k$ minimal models 
are extensively discussed in the review by Bouwknegt and Schoutens \cite{w_k};
we are going to use their formula (6.50) 
for the spectrum of the ${\cal W}_k$ operators derived by the 
quantum Miura transformation \cite{w_k}:
\beq\label{spec}
{\widetilde{\cal W}}_a(\L)=(-1)^{a-1} \sum_{1\leq i_1<i_2< \cdots < i_{a} 
\leq k }
\quad \prod_{s=1}^a \left[ (\L | \h_{i_s}) + (a-s) \alpha_0 \right]\ ,
\eeq
with $a=2,\dots , k$.
 We should stress  
that the fields ${\widetilde{\cal W}}_a(z)$ are {\it not} quasi-primary 
and therefore are not exactly
proportional to the Casimir operators of the coset (except for the field 
$\widetilde{\cal W}_2 ={\cal W}_2$, the stress tensor); nevertheless, 
the Casimir operators are linear combinations of the fields 
$\widetilde{\cal W}_a$ and their 
derivatives. For example, the third order Casimir field  
in the  ${\cal W}_k$ algebra is given by:
\[
{{\cal W}_3(z)} = {\widetilde{\cal W}_3(z)} - \frac{k-2}{2} \alpha_0 
\partial_z {\widetilde{\cal W}_2(z)}.
\]

The Coulomb gas parameters in Eq. (\ref{spec})  are given by:
\[
\alpha_+ + \alpha_-=\alpha_0, \quad  \alpha_+ \alpha_-=-1 \quad 
\mathrm{and} \quad
(\alpha_0)^2=\frac{1}{p(p+1)} \quad  \mathrm{for} \quad p=k+1.
\]
The vectors $\{\h_i \}_{i=1}^{k}$ form the standard over-complete 
basis for the $su(k)$ root lattice; they satisfy:
\beq\label{h_i}
\sum_{i=1}^k \h_i=0,
\eeq
and express the positive roots and the fundamental weights by
\beq
\label{roots}
\a^i = \h_i-\h_{i+1}\ ,\qquad
\L_i = \sum_{s=1}^i \h_s\ , \qquad i=1,\ldots, k-1; 
\eeq
therefore:
\beq\label{h-metrics}
\h_i=\L_i - \L_{i-1}, \quad i=1,\ldots, k, \qquad
(\h_i|\h_j)=\delta_{ij} - \frac{1}{k}\ ;
\eeq
(assuming  $\L_k\equiv \L_0 \equiv 0$). In particular, the Weyl vector
has two different representations (the $\h_i$ are linearly dependent):
\[
\underline{\rho}=\sum_{i=1}^{k-1} \L_i = \frac{1}{2}
\sum_{i=1}^{k}  \left( k+1-2i\right) \h_i =
 -\sum_{i=1}^{k}  i \; \h_i.
\]
In order to compute the eigenvalues of the  ${\cal W}_k$ generators 
we  make the following identification of the weight:
\beq\label{weight}
\L\equiv \alpha_- (\L_\mu +\L_\nu ) , \quad \mathrm{where} \quad
\alpha_-=- \frac{p}{\sqrt{p(p+1)}}\  .
\eeq
Using these formulas, we obtain the quantum numbers 
of the irreducible representations of the  $\widetilde{\cal W}_3$ and 
$\widetilde{\cal W}_4$ algebras,
which are listed in the Tables \ref{tab.3} and \ref{tab.w_4},
respectively.

{\bf Remark.} 
In general, the Fateev--Lukyanov weights $\beta$ \cite{w_k} are expressed
as $\beta=\alpha_+ \L + \alpha_- \L'$ where $\L$  and $\L'$ 
are level 1 and 2 $su(k)$ weights respectively. 
From the coset point of view, however, the irreducible representations 
are labeled by  triples
consisting of two level 1 weights  $\L_\alpha$, $\L_\beta$  and one 
level 2 weight $\L$ (see Eq. (\ref{3.1})). 
According to Eq. (\ref{3.3}) the level 2 weight  $\L$
together with one of the level 1 weights determine the entire triple 
since the remaining level 1 weight is fixed by charge conservation.
It does not matter which of the level 1 weights we choose,
so we characterize the triple by the second and the 
third weight ($\L_\beta$, $\L$).
In our case ($p=k+1$), we can use the
symmetries (\ref{3.4}) of the pairs $(\L_\beta,\L)$ to put the 
level-one weight to zero as in Eq. (\ref{3.6}); furthermore,
we can always write a level-two $su(k)$ weight as a 
sum of two fundamental weights. 
In conclusion, the Fateev--Lukyanov weights are given by (\ref{weight})
(up to the factor $\sqrt{2}$ for any $a$ adjusting the normalization).

\begin{table}
\begin{center}
\begin{tabular}{| c  || c |  c | c | c |}
\hline
Label & Field  & Dimension $\D$ & $\Z_3$-charge $p$ & 
$\widetilde{\cal W}_3$ charge \\
\hline\hline
$\L_0 +\L_0$ & $1$ & $0$ & $0$ & $0$ \\
\hline
$\L_0 +\L_1$ & $\s_1$ & $\frac{1}{15}$ & $1$ & $-\frac{7}{675}\sqrt{5}$ \\
\hline
$\L_0 +\L_2$ & $\s_2$ & $\frac{1}{15}$ & $2$ & $-\frac{2}{675}\sqrt{5}$ \\
\hline
$\L_1 +\L_1$ & $\psi_1$ & $\frac{2}{3}$ & $2$ & $-\frac{22}{135}\sqrt{5}$ \\
\hline
$\L_1 +\L_2$ & $\varepsilon$ & $\frac{2}{5}$ & $0$ & $-\frac{1}{25}\sqrt{5}$ \\
\hline
$\L_2 +\L_2$ & $\psi_2$ & $\frac{2}{3}$ & $1$ & $\frac{4}{135}\sqrt{5}$ \\
\hline
\end{tabular}
\end{center}
\caption{Primary fields of the $\Z_3$-parafermion theory and their 
quantum numbers, together with the eigenvalues of the 
$\widetilde{\cal W}_2$ and $\widetilde{\cal W}_3$
currents and of the $\Z_3$ charge.} 
\label{tab.3}
\end{table}

\begin{table}[t]
\begin{center}
\begin{tabular}{|c||c|c|c|}
\hline
${\L _{\mu}} + {\L _{\nu}}$ & $\Delta(\L_\mu+\L_\nu)$ & 
$\widetilde{\cal W}_3(\L_\mu+\L_\nu)$ &
$\widetilde{\cal W}_4(\L_\mu+\L_\nu)$ \cr\hline\hline
$ \L _{0} + \L_0 $ & $0$ & $0$ &$0$ \cr \hline
${\L _{0}} + {\L _{1}}$ & $\frac {1}{16}$ & $- \frac {1}{160} \,\sqrt{30}$ &
$-\frac {3}{1280} \,\sqrt{30} $ \cr \hline
${\L _{0}} + {\L _{2}}$ & $\frac {1}{12}$ & $- \frac {1}{180} \,\sqrt{30}$ &
$\frac {7}{1440} \,\sqrt{30}$ \cr \hline
${\L _{0}} + {\L _{3}}$ & $\frac {1}{16}$ & $- \frac {1}{480} \,\sqrt{30}$ &
$-\frac {7}{3840} \,\sqrt{30}$ \cr \hline
$\L_{1}+\L_1$ & $\frac {3}{4}$ & $- \frac {7}{60} \,\sqrt{30}$ &
 $- \frac {7}{160} \,\sqrt{30}$ \cr \hline
${\L _{1}} + {\L _{2}}$ & $\frac {9}{16}$ & $- \frac {29}{480}\,\sqrt{30}$ &
$-\frac {247}{3840}\,\sqrt{30}$ \cr \hline
${\L _{1}} + {\L _{3}}$ & $ \frac {1}{3}$ & $-  \frac {1}{45} \,\sqrt{30}$ &
$-\frac {1}{90} \,\sqrt{30}
$ \cr \hline
$\L_{2}+\L_2$ & $1$ & $- \frac {1}{15} \,\sqrt{30}$ &
$ - \frac {1}{15} \,\sqrt{30} $ \cr \hline
${\L _{2}} + {\L _{3}}$ & $\frac {9}{16}$ & $ - \frac {7}{480} \,\sqrt{30}$ &
$- \frac {3}{256} \,\sqrt{30}$ \cr \hline
$\L _{3}+ \L_3$ & $\frac {3}{4}$ & $\frac {1}{60} \,\sqrt{30}$ &
$ - \frac {1}{160} \,\sqrt{30}
$ \cr \hline
\end{tabular}
\end{center}
\caption{$\Z_4$-parafermion theory: labels of irreducible representations
and eigenvalues of the
$\widetilde{\cal W}_2$,  $\widetilde{\cal W}_3$  and $\widetilde{\cal W}_4$ 
currents.} 
\label{tab.w_4}
\end{table}


\sectionnew{Example: the $\Z_3$-parafermion Hall state}

We now spell out the coset construction of Section 4 in the case of $k=3$
(the analysis of the Pfaffian state ($k=2)$ can be found in
Ref.\cite{cgt}).
Consider the lattice model of Section 3 for
$k=3$, $M=1$: the  coset (\ref{1.4}) coincides in this case with
the $\Z_3$ Potts model, with $c_\PF=4/5$, which is also
the well-known third minimal model of the Virasoro algebra;
this model possesses six 
sectors, corresponding to the irreducible representations
of the ${\cal W}_3$ algebra. On the other hand, 
the complete theory $\PF_3\oplus\U1$ for the parafermion Hall state  
has ten sectors (this is the value of the topological order (\ref{3.23})).
Let us discuss both theories in sequence.

The labels $(\L_\mu +\L_\rho)$ of the six parafermion
primary fields given by the coset construction, their
conformal dimensions
$\D(\L_\mu +\L_\rho)$, the $\Z_3$ charges $p$ and the ${\cal W}_3$ 
charges are given in Table~\ref{tab.3} .
We can label the six sectors
by an integer $\lambda$,  $-2 \leq \lambda \leq 3$,  related to the pair
$(\mu,\rho)$  ($0\leq \mu,\rho \leq 2$) of Eq. (\ref{3.7}) as follows:
$0=(0,0)$, $1=(1,0)=(0,1)$, $ \; -1=(2,0)=(0,2)$, $\; 2=(1,1)$, $\; -2=(2,2)$,
$\; 3=(1,2)$.
The conformal dimensions $\D_\lambda$ are given by:
\beq\label{3.5'}
\D_0=0, \quad \D_{\pm 1}=\frac{1}{15}, \quad \D_{\pm 2}=\frac{2}{3},
\quad  \D_{ 3}=\frac{2}{5}.
\eeq
The six coset characters (\ref{3.16}) for $k=3$ 
(hereafter also called $ch_\lambda$) are:
\beqa
\label{4.9}
\!\!\!\!\! ch_{2l}(\t)&\!\! =\!\! &\ch_{0,l}(\t) = q^{-\frac{1}{30}}
\!\!\!\!\! \sum_{n_1+2n_2=l \mod 3,\ \ (n_i\geq 0)}
\frac{q^{\frac{2}{3}\left(n_1^2+n_2n_2+n_2^2 \right)}}{(q)_{n_1}(q)_{n_2}},
\quad l=0,\pm 1 (\mod 3) \ ,\nn
\!\!\!\!\! ch_{2l-1}(\t)&\!\! =\!\! &  \ch_{l+1,l+1}(\t) =
q^{\frac{1}{30}} \!\!\!  \sum_{n_1+2n_2=0 \mod 3}
\frac{q^{\frac{2}{3}\left(n_1^2+n_2n_2+n_2^2 \right) -
\frac{2n_{l+1}+n_{2-l}}{3}}}{(q)_{n_1}(q)_{n_2}}, \quad l=0, 1\ , \nn
\!\!\!\!\! ch_{3}(\t)&\!\! =\!\! & \ch_{1,2}(\t) =
q^{\frac{1}{30}} \!\!\! \sum_{n_1+2n_2=2 \mod 3}
\frac{q^{\frac{1}{3}\left[2n_1(n_1-1)+2 n_1 n_2 + n_2(2n_2-1) \right]}}
{(q)_{n_1}(q)_{n_2}}.
\eeqa
These match the Rocha-Caridi characters 
$\chi_{\D}$ of the $c=4/5$ Virasoro minimal model \cite{rc}:
\beqa
\label{ident}
\ch_{0,0}&=&\chi_{0} + \chi _{3}\ , \nn
\ch_{0,1}=\ch_{0,2} &=&\chi_{2/3}\ , \nn
\ch_{1,0}=\ch_{2,0} &=&\chi_{1/15}\ , \nn
\ch_{1,2} &=&\chi_{2/5} + \chi _{7/5}\ .
\eeqa
The Potts-model partition function on the torus is given by the sum
of the absolute value squared of these characters;
these results verify that the symmetry algebra of this model is 
Virasoro extended by the ${\cal W}_3$ algebra \cite{w_k}.

The fusion rules of the primary fields in the theory are
derived with the help of Section 4.4 and are listed in Table \ref{tab-f}
(See Table \ref{tab.3} for the identification of the fields with their
coset labels).
Using these rules, it was found that the number of conformal blocks
produced in the fusion of $3m$ spin fields $\s_1$ is equal to the
Fibonacci number $F_{3m-2}$ \cite{rr}.
Here we are interested in checking the assignment of the $\Z_3$ 
parafermion ``charge'' $p$ and its interplay with the fractional part of the
physical charge ($\U1$ part) in the full theory discussed hereafter.

The $\Z_3$-parafermion Hall state is characterized by the chiral algebra 
$\A_3$ defined by (\ref{3.11}), the filling fraction $\nu_3=3/5$  and 
the central charge $c=9/5$; its ten sectors
are described by the following characters:
\beqa
\label{3.8'}
\chi_{m \mu}(\t,\z)&=&\sum_{l=-1}^1 K_{m+5l} (\t,3\z;15) ch_{m+3\mu+2l}(\t)
\ , \nn
&& m=0,\pm 1, \pm 2, \quad \mu=0,1\ , 
\qquad\quad (ch_\lambda(\t)=ch_{\lambda+6}(\t) ) \ .
\eeqa
which are obtained by combining
the charged ($K$) and neutral ($ch$) characters (see Eqs. (\ref{3.12}) and 
(\ref{4.9})) according to the $\Z_3$ parity rule (\ref{p-rule}).

The minimal charges $Q_{m \mu}$ and minimal conformal
dimensions $\D_{m \mu}$ of Hall edge excitations 
in the sectors (\ref{3.8'}) are:
\beqa
\label{4.13}
 Q_{m, \mu}&=& \frac{m}{5}\ , \qquad \qquad  m=0,\pm 1, \pm 2, \quad \mu=0,1,
\nl
\D_{m, 0} &=& \frac{1}{10}{|m|+1 \choose 2},\qquad
\D_{0, 1}=\frac{2}{5}=2\D_{\pm 2, 1}, \qquad \D_{\pm 1,1}=\frac{7}{10}.
\eeqa

\begin{table}
\begin{center}
\begin{tabular}{lllll}
$\s_1\cdot \s_1 =  \s_2 + \psi_1$ & & $\s_2\cdot \s_2 =\s_1+\psi_2$ & &
$\psi_1 \cdot\psi_1=\psi_2$
\\
$\s_1\cdot \s_2 = 1 + \ep$        & & $\s_2\cdot \psi_1 =  \s_1$    & &
$\psi_1\cdot \ep =  \s_2$
\\
$\s_1\cdot \psi_1 = \ep$         & & $\s_2\cdot \ep =  \s_2 + \psi_1$ & &
$\psi_1\cdot \psi_2 = 1 $
\\
$\s_1\cdot \ep = \s_1 + \psi_2$  & & $\s_2\cdot \psi_2 =  \ep $      & &
$\ep \cdot \ep = 1+ \ep$
  \\
$\s_1\cdot \psi_2 = \s_2$       & &  $\psi_2\cdot \psi_2=\psi_1$  & &
$\ep \cdot \psi_2 = \s_1$ 
\end{tabular}
\end{center}
\caption{Fusion rules of the $\Z_3$-parafermion theory.}
\label{tab-f}
\end{table}

Quantum Hall states in the geometry of an annulus can be characterized
by a (real) partition function which is (weakly) modular invariant, 
as described in Ref.\cite{cz}; this quantity completely
accounts for the edge excitations on the two edges, which carry opposite 
chirality and satisfy some global conditions, such as total integer charge 
and half-integer spin.
The partition function for the $\Z_3$-parafermion Hall state
is given by the diagonal modular invariant:
\beq
Z(\tau,\zeta)= \sum_{m,\mu}\ \left\vert \widetilde\chi_{m\mu} 
\right\vert^2 \ ;
\label{zed}
\eeq
(this form obviously extends to higher $k$ values). In this equation,
the characters $\widetilde\chi$ are given by the complete
characters (\ref{3.8'}) times a convenient real factor \cite{cz}:
$\widetilde\chi_{m\mu}=\exp\left( 
- {\pi\over 5}{(\I{\zeta} )^2\over \I{\tau}} \right) \chi_{m\mu}$.
The invariance of the partition function under the modular
transformations $T^2:\ (\tau,\zeta)\to(\tau+2,\zeta)$ and 
$S:\ (\tau,\zeta) \to \left( -1/\tau,\zeta/\tau \right)$
holds because the characters (\ref{3.8'}) transform linearly
by unitary matrices. In particular, the $S$ modular inversion 
is represented by the matrix (up to an overall phase \cite{cz}):
\beq\label{3.10'}
S_{m\mu,m'\mu'}=\frac{\sqrt{3-\d}}{5}\ 
(-1)^{\mu\mu' +m\mu'+m'\mu}\ \ex^{i\frac{\pi}{5} m m'}\ 
\d^{P_{\mu+m+\mu'+m'}} \ ,
\eeq
with $m,m'=0,\pm1,\pm 2$,  $\mu,\mu'=0,1$,
$\ex^{i\pi/5}=(\d+i\sqrt{3-\d})/2$ and
$P_{\lambda} = \left( 1-(-1)^{\lambda}\right)/2$.

The matrix elements of $S$ determine the so-called
quantum dimension $d_{q}$ of the representation $(m,\mu)$, which is:
\beqa\label{3.11'}
d_q(m,\mu)= \frac{S_{m \mu,00}}{S_{00,00}}=
\left\{ \begin{array}{rll} 1 & \mathrm{for} \; \mu+m & \mathrm{even} , \\
\d & \mathrm{for} \; \mu+m & \mathrm{odd}\ . \end{array} \right.
\eeqa
The presence of a non-integer (in fact, irrational) quantum dimension
$\d$ signals the non-abelian statistics of the quasi-particle
excitations in this Hall state (already discussed at the
level of wave functions in Section 2.1). Indeed, some fusion
rules yields more than one field in the r.h.s., as shown in Table \ref{tab-f}.


\sectionnew{Conclusions and a general framework}

In this paper, we have described
 the conformal theory of $\Z_k$-parafermion Hall states by a two-step
procedure: first introducing a lattice current 
algebra theory corresponding to the same filling fraction, and then
taking a coset projection. 
We have presented two abelian theories, with central charge
$c=k$ and $c=2k-1$, respectively. The $c=k$ one  is useful
to write wave functions which nicely describe the origin
of the non-abelian statistics of quasi-particles as the effect of 
the projection.
The $c=2k-1$ theory accounts for a complete definition
and description of all excitations of the parafermion
Hall states, which are deduced from a highly
symmetric choice of charge lattice followed by the coset construction.

In conclusion we would like to briefly outline a
general program of defining and classifying
conformal field theories suitable for describing quantum Hall states.
Our objective here would be to directly characterize the outcome - 
without reference to the parent abelian theory - and find the
lattice theories and coset models as special cases,
in the same spirit as in Ref.\cite{fpsw}.
We sum up the basic characteristics of the theory by a set of 
four postulates:

{\bf P1}. Each Hall plateau and its excitations are described by a 
rational conformal field theory 
with a local chiral algebra ${\cal A}$ which contains a $\U1$ "electric 
current" $J(z)$ and an odd (Fermi) electron field $\Psi_e$ of charge 
$(\Q |\q_e)=-1$  and minimal (in ${\cal A}$) half-integer 
dimension $\D_e=p+1/2$ (and its conjugate field $\Psi_e^*$).

{\bf P2}. The neutral sub-algebra ${\cal A}_0$ of ${\cal A}$,
\beq
{\cal A}_0=\U1\otimes {\cal C}\ , \qquad
\left( \Phi_e(z) \in {\cal C}\quad {\rm iff}\quad
[J(z), \Phi_e(w)]=0  \right)\ ,
\label{c1}   
\eeq
is bosonic. The electron is associated to a primary (with respect 
to ${\cal A}_0$) - field $\Psi_e$ satisfying the fusion rule:
\beq
\Psi_e^* \cdot \Psi_e \ \in {\cal A}_0\ .
\label{c2}
\eeq
(Hence, it is a "simple current", see  Section 5.1 of \cite{fpsw}.)

{\bf P3}. The electron field transforms under a tensor product representation 
of $\U1\otimes {\cal C}$:
$\Psi_e(z)=Y((\nu_k)^{-1/2}, z)\ \phi_k(z)$, such that:
\beq 
\phi_k(z_1)\cdots \phi_k(z_k)\ \in \ {\cal A}_0 \ ;                  
\label{c3}
\eeq
here, $k$ is a multiple of the numerator of the fractional part
$\nu_k$ of the filling fraction ($0<\nu_k<1$). 
Its conformal dimension splits correspondingly into a sum of two terms
(cf. Eq.(\ref{dim})):
\beq
2 \D_e={1\over\nu_k} + 2 \D_0 \ .
\label{c4}
\eeq
In particular, for a Laughlin fluid, $k=1$ and $\phi_1(z_1)=1$.
This postulate represents the $\Z_k$ selection rule in the general 
framework.

{\bf P4}. The admissible excitations (or the superselection sectors) of the 
Hall state are described by the primary fields of ${\cal A}$ 
that are, in particular, relatively local to $\Psi_e$.         
(This is, in fact, part of the definition of a rational conformal
field theory with chiral algebra ${\cal A}$.)

Here are some implications of these postulates:

i) The rationality of the theory implies that the filling fraction and
all conformal dimensions are rational numbers.

ii) The charge-statistics relation (which generalizes (\ref{0.3b}): 
odd and even square charges correspond to Fermi and Bose fields,
respectively) is a consequence of P1 and P2. 
Indeed, (composite) Bose fields in ${\cal A}_0$ contain an even number of
electron field factors and, hence, carry an even charge.

iii) If $\nu_k=n/d$ (where $n, d$ are natural numbers) then 
$k=l n$ and
the minimal (non zero) absolute value of the charge of an excitation
is $1/(l d)$ (cf. Section 5.1 of \cite{fpsw}).

iv) The characters of the superselection sectors of the conformal
theory with chiral 
algebra ${\cal A}$ (obeying P4) span a representation of the subgroup 
$\G_2$ of the modular group $SL(2, \Z)$ generated by $S$ and 
$T^2$.
If we adopt, in addition, the postulates (or, rather, observations) 
(S1) and (S3) of \cite{fpsw}, saying that a Hall state is more stable the 
smaller is the Virasoro central charge and the topological order
(i.e., the number of independent excitations), then the coset models 
considered in this paper are clearly favoured compared to the 
lattice theories describing the same filling fractions.

\bigskip
{\bf Acknowledgements}

L. G. thanks C. Schweigert for fruitful discussions.
L. G. thanks I.N.F.N., Firenze for hospitality;
I. T. thanks  the Institut des
Hautes Etudes Scientifiques (IHES), Bures-sur-Yvette, for hospitality.
L.G. and I.T. acknowledge partial support from the Bulgarian
National Foundation for Scientific Research under contract F-828.
A.C. acknowledges the partial support from the European
Community network programme FMRX-CT96-0012.

\def\NP{{\it Nucl. Phys.\ }}
\def\PRL{{\it Phys. Rev. Lett.\ }}
\def\PL{{\it Phys. Lett.\ }}
\def\PR{{\it Phys. Rev.\ }}
\def\CMP{{\it Commun. Math. Phys.\ }}
\def\IJMP{{\it Int. J. Mod. Phys.\ }}
\def\JSP{{\it J. Stat. Phys.\ }}
\def\JP{{\it J. Phys.\ }}

\end{document}